\pdfoutput=1

\documentclass[11pt,twoside,a4paper,cmspaper,final,collab]{cms-tdr}
\def\svnVersion{b04ca73-D}\def\svnDate{2020/01/14}\def\cmsCernNoTag{CERN-EP-2019-192}\def\cmsCernDate{\today}\def\cmsMessage{"Published in the Journal of High Energy Physics as \href{http://dx.doi.org/10.1007/JHEP02(2020)015}{\doi{10.1007/JHEP02(2020)015}.}"}

\begin{document}\cmsNoteHeader{SUS-19-003}

\hyphenation{had-ron-i-za-tion}
\hyphenation{cal-or-i-me-ter}
\hyphenation{de-vices}
\newcommand{\emu}{{\Pe}\PGm}
\newcommand{\mumu}{{\PGm}\PGm}
\newcommand{\tauhtauh}{{\tauh}\tauh}
\newcommand{\PSGtpDo}{\HepSusyParticle{\PGt}{1}{+}\Xspace}
\newcommand{\PSGtpmDo}{\HepSusyParticle{\PGt}{1}{\pm}\Xspace}
\newcommand{\cmsTable}[1]{\resizebox{\textwidth}{!}{#1}}
\newcommand{\SYSCALC} {\textsc{SysCalc}\xspace}

\newlength\cmsTabSkip\setlength{\cmsTabSkip}{1ex}
\newlength\cmsTabSkipLarge\setlength{\cmsTabSkipLarge}{4ex}

\hyphenation{had-ron-i-za-tion}
\hyphenation{cal-or-i-me-ter}
\hyphenation{de-vices}

\cmsNoteHeader{SUS-19-003}
\title{Search for top squark pair production in a final state with two tau leptons in proton-proton collisions at $ \sqrt{s} = 13\TeV $}

\date{\today}

\abstract{A search for pair production of the supersymmetric partner of the top quark, the top squark, in proton-proton collision events at $ \sqrt{s} = 13\TeV $ is presented in a final state containing hadronically decaying tau leptons and large missing transverse momentum. This final state is highly sensitive to high-\tanb or higgsino-like scenarios in which decays of electroweak gauginos to tau leptons are dominant. The search uses a data set corresponding to an integrated luminosity of 77.2\fbinv, which was recorded with the CMS detector during 2016 and 2017. No significant excess is observed with respect to the background prediction. Exclusion limits at 95\% confidence level are presented in the top squark and lightest neutralino mass plane within the framework of simplified models, in which top squark masses up to 1100\GeV are excluded for a nearly massless neutralino.}

\hypersetup{%
pdfauthor={CMS Collaboration},%
pdftitle={Search for top squark pair production in a final state with two tau leptons in proton-proton collisions at sqrt(s) = 13 TeV},%
pdfsubject={CMS},%
pdfkeywords={CMS, supersymmetry, top quark}}

\maketitle

\section{Introduction}

Supersymmetry (SUSY)~\cite{Ramond:1971gb,
	Golfand:1971iw, Neveu:1971rx, Wess:1973kz, Fayet:1974pd,
	tHooft:1979rat, Kaul:1981hi, Nilles:1983ge, Martin:1997ns}
is one of the most widely studied theories of physics
beyond the standard model (SM),
providing solutions to various
shortcomings of the SM.
In SUSY models there is a bosonic superpartner for each fermion (and
vice-versa), the superpartner having the same quantum numbers, other
than spin, as its
SM partner.
The superpartners of the SM gauge and Higgs bosons (gauginos
and higgsinos, respectively) mix to produce charginos and neutralinos.
The weakly interacting lightest neutralino \PSGczDo can be a dark
matter candidate in $R$-parity conserving SUSY models~\cite{Farrar:1978xj}.
The SUSY partners of left- and right-handed top quarks are the top squarks,
\stL and \stR. These particles can mix with each other, resulting in physical states
\stone and \sttwo, with \stone defined to be the lighter of the two.
The top squarks play an important role in stabilizing the Higgs boson mass by canceling the dominant top quark loop correction.
Therefore, there is a strong motivation to perform searches for top squark production.

In this study, we focus on the signal of top squark pair production in a final state with two tau leptons.
This probes the part of the parameter space of the minimal
supersymmetric standard model (MSSM) in which the lightest
charginos (\PSGcpmDo) and neutralino preferentially couple to third-generation fermions, such as tau leptons.
The interaction of the charginos and neutralinos with fermion-sfermion
pairs involves both gauge and Yukawa terms~\cite{Martin:1997ns}, so if
charginos and neutralinos are predominantly higgsino-like, they will
preferentially couple to third-generation fermion-sfermion pairs
through the large Yukawa coupling.
Moreover, the Yukawa coupling to the tau lepton-slepton pairs can be large for a high value of \tanb
even if the higgsino component is relatively small.
Additionally, a large value of \tanb can make the lighter state of the
superpartner of the tau lepton ($\PSGtDo$) much lighter than
the superpartners of the first and second generation leptons.
Consequently, the chargino decays predominantly
as $ \PSGcpDo \to \PSGtpDo \PGnGt $ or $ \PGtp \PSGnGt $ (charge conjugation is assumed throughout in this paper),
and the decay rates in the electron
and muon channels are greatly reduced~\cite{Baer:1997yi, Guchait:2002xh}.
Therefore, searches for SUSY signals in electron and muon channels are less sensitive to this scenario.

We focus on the top squark decays
$\PSQtDo \to \PQb \PSGcpDo \to \PQb \PSGtpDo \PGnGt \to  \PQb \PGtp \PSGczDo \PGnGt$
~and~
$\PSQtDo \to \PQb \PSGcpDo \to \PQb \PGtp \PSGnGt \to \PQb \PGtp \PSGczDo \PGnGt$.
The \PSGczDo is assumed to be
the lightest SUSY particle (LSP). Being neutral and weakly interacting, it
leaves no signature in the detector, resulting in an imbalance
in transverse momentum \pt.
The neutrinos produced in the decay chains also contribute to the \pt imbalance.
Hence, the events of interest contain two tau leptons,
two {\PQb} quarks, and a \pt imbalance.
The decay chains are depicted by the four diagrams in Fig.~\ref{fig:sigFeyn} within the simplified model spectra (SMS)
framework~\cite{Alwall:2008ag, Alves:2011wf}.
It is assumed that the $\PSGcpDo$ decays to \PSGtpDo or \PSGnGt with equal probability.

\begin{figure}[!htbp]
	\centering
	
	\includegraphics[width=0.45\textwidth]{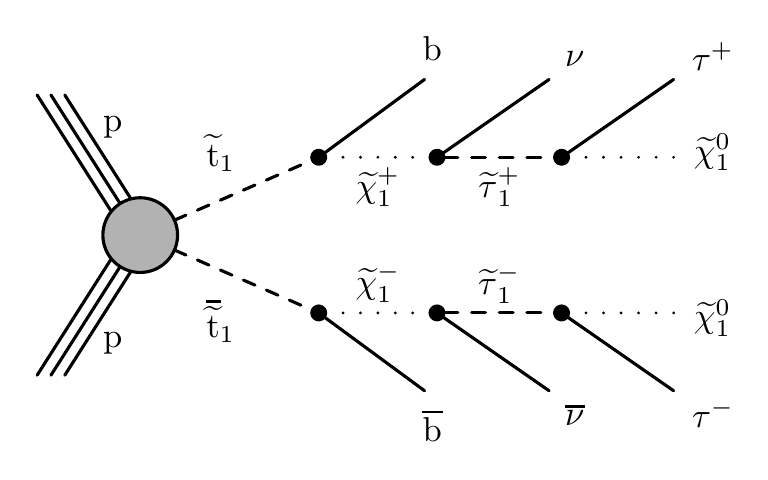}
	\includegraphics[width=0.45\textwidth]{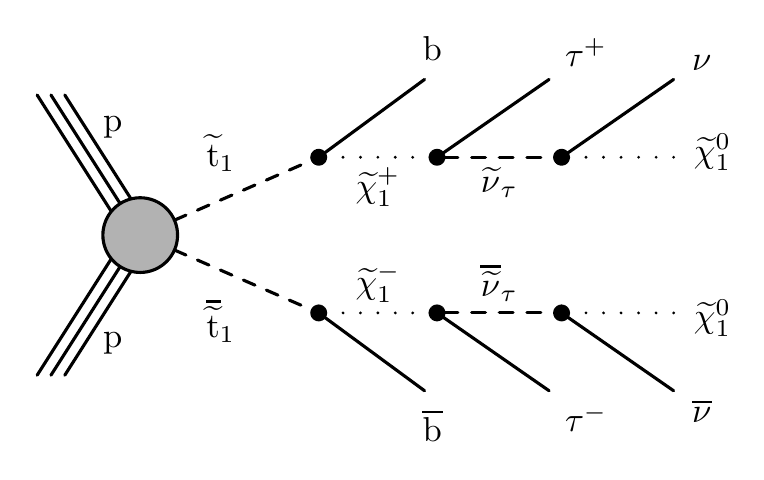} \\
	\includegraphics[width=0.45\textwidth]{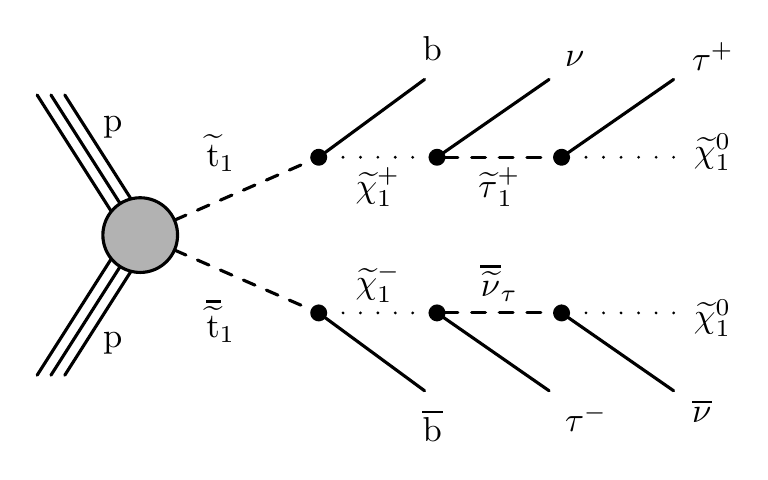}
	\includegraphics[width=0.45\textwidth]{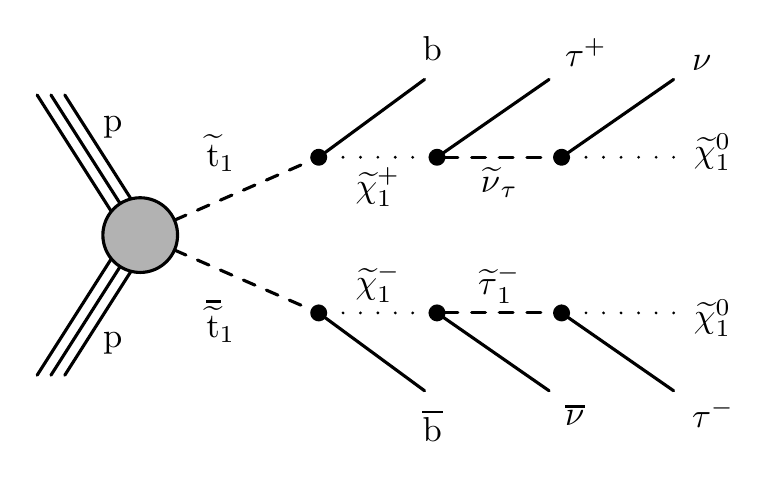}
	
	\setcounter{figure}{0}
	\caption{Top squark pair production in proton-proton collisions at the LHC, producing pairs of \PQb quarks and taus accompanied by neutrinos and LSPs in the final state.
	}
	\label{fig:sigFeyn}
\end{figure}

This search is performed using proton-proton collision events at a
center-of-mass energy of 13\TeV, recorded by the CMS experiment at the CERN LHC.
The data sample corresponds to integrated luminosities of
35.9 and 41.3\fbinv
collected during the 2016 and 2017 operating periods of the LHC, respectively.
Signal-like events are characterized by the presence of hadronically
decaying tau leptons (\tauh), jets identified as likely to have
originated from the fragmentation of \PQb quarks, and large missing \pt.
Contributions from SM processes with the same final state are estimated using a combination of Monte Carlo (MC) simulated samples and control samples in data.

Searches for top squark pair production in leptonic and hadronic final states have been performed by the
CMS \cite{Sirunyan:2017xse, Sirunyan:2017leh, Chatrchyan:2013xna, Khachatryan:2016pup, Khachatryan:2016pxa, Sirunyan:2016jpr, Sirunyan:2017wif, Sirunyan:2017pjw} and ATLAS \cite{Aaboud:2017nfd, Aad:2015pfx, Aad:2014kra, Aad:2014qaa, Aaboud:2016lwz} Collaborations, establishing limits on top squark masses in the framework of SMS models. The ATLAS Collaboration performed a search~\cite{PhysRevD.98.032008} based on 2015 and 2016 data probing the same final state as that used here, but optimized for a
gauge-mediated SUSY breaking scenario with an almost massless gravitino as a source of missing momentum. Therefore, final states containing hadronically decaying tau leptons have not been extensively explored in the context of top squark searches motivated by high-\tanb and higgsino-like scenarios.

The paper is organized as follows. A brief description of the CMS detector is presented in Section \ref{sec:CMSdetector}, followed by
descriptions of the event simulation in Section~\ref{sec:MCsimulation}, and
reconstruction in Section~\ref{sec:EventReco}.
The event selection and search strategy are detailed in
Section~\ref{sec:EventSel}.
We explain the various methods used for background estimation in
Section~\ref{sec:BkgEstimation}, the systematic uncertainties are discussed in Section \ref{sec:Systematics}, and the results are provided in
Section~\ref{sec:Results}.
Finally, the analysis is summarized in Section~\ref{sec:Summary}.

\section{The CMS detector}
\label{sec:CMSdetector}

The central feature of the CMS apparatus is a superconducting solenoid of 6\unit{m} internal diameter, providing a magnetic field of 3.8\unit{T}. Within the solenoid volume are a silicon pixel and strip tracker, a lead tungstate crystal electromagnetic calorimeter (ECAL), and a brass and scintillator hadron calorimeter (HCAL), each composed of a barrel and two endcap sections. Forward calorimeters extend the pseudorapidity coverage provided by the barrel and endcap detectors. Muons are detected in gas-ionization chambers embedded in the steel flux-return yoke outside the solenoid.
A more detailed description of the CMS detector, together with a definition of the coordinate system used and the relevant kinematic variables, can be found in Ref.~\cite{Chatrchyan:2008zzk}.

Events of interest are selected using a two-tiered trigger system~\cite{Khachatryan:2016bia}. The first level, composed of custom hardware processors, uses information from the calorimeters and muon detectors to select events at a rate of around 100\unit{kHz} within a time interval of less than 4\mus. The second level, known as the high-level trigger, consists of a farm of processors running a version of the full event reconstruction software optimized for fast processing, and reduces the event rate to around 1\unit{kHz} before data storage.

\section{Monte Carlo simulation}
\label{sec:MCsimulation}

Simulation is used to estimate several of the SM backgrounds.
The predictions for signal event rates are also estimated using
simulation, based on simplified SUSY signal models.
The simulation is corrected for small discrepancies observed with
respect to collision data using a number of scale factors (SFs).
These will be discussed in later sections.

The pair production of top quarks (\ttbar) is generated at
next-to-leading order (NLO) in \alpS using \POWHEG v2~\cite{OLEARI201036, Nason:2004rx, Frixione:2007vw, Alioli:2010xd, Frixione:2007nw}.
The same \POWHEG generator has been used for the single top quark $t$-channel
process, whereas \POWHEG v1 has been used for the tW process~\cite{Alioli:2009je}.
The \MGvATNLO v2.2.2 (v2.4.2 for 2017)~\cite{Alwall:2014hca} generator is used at leading order (LO) for modeling the Drell--Yan+jets (DY+jets) and W+jets backgrounds, which are normalized to the next-to-next-to-leading order (NNLO) cross sections.
The \MGvATNLO generator is also used at NLO for simulating the
diboson, $ \PV\PH $, and $ \ttbar\PV $ (\PV = \PW or \PZ) backgrounds.
For the 2016 analysis, the parton shower and hadronization are simulated with \PYTHIA v8.212~\cite{Sjostrand:2014zea} using the underlying event
tunes CUETP8M2T4~\cite{CMS-PAS-TOP-16-021} (for \ttbar only) or CUETP8M1~\cite{Khachatryan:2110213}.
For the 2017 analysis, \PYTHIA v8.230 with the tune
CP5~\cite{Sirunyan:2019dfx} is used.
The CMS detector response is modeled using \GEANTfour
\cite{Agostinelli:2002hh}, and the simulated events are then reconstructed in the same way as collision data.

Signal processes for top squark pair production shown in
Fig.~\ref{fig:sigFeyn} are generated at LO using \MGvATNLO v2.2.2.
The tunes CUETP8M1 and CP2~\cite{Sirunyan:2019dfx} are used for the
2016 and 2017 analyses, respectively.
The signal cross sections are evaluated using NNLO plus
next-to-leading logarithmic (NLL)
calculations~\cite{bib-nlo-nll-01, bib-nlo-nll-02, bib-nlo-nll-03, bib-nlo-nll-04, bib-nlo-nll-05}.
Detector response for the signal events is simulated using the fast CMS
detector simulation (\textsc{FastSim})~\cite{Giammanco:2014bza}.

We assume a branching fraction  of 50\%  for each of the two decay modes of the chargino, $ \PSGcpDo \to \PSGtpDo\PGnGt$ and $ \PSGcpDo \to \PGtp\PSGnGt$. Each of the four diagrams in
Fig.~\ref{fig:sigFeyn} therefore contributes 25\% of the generated signal events.
The masses of SUSY particles appearing in the decay chain are determined by
the parameterization
\begin{equation}
\begin{aligned}
m_{\PSGcmDo} - m_{\PSGczDo} &= 0.5 \
(m_{\stone} - m_{\PSGczDo} ) ,
\\
m_{\PSGt_{1}} - m_{\PSGczDo} &= x \
(m_{\PSGcmDo} - m_{\PSGczDo} ) ,
\\
x &\in [0.25,\ 0.5,\ 0.75] ,
\\
m_{\PSGnGt} & = m_{\PSGt_{1}} .
\end{aligned}
\label{eq:mass}
\end{equation}
In this parameterization, the chargino mass is fixed to be the mean of
the top squark and $\PSGczDo$ masses. The masses of the leptonic
superpartners are set by the value of $ x $ for a given pair of top
squark and $\PSGczDo$ masses. The kinematic properties of the
final state particles in each of the decay chains depicted in
Fig.~\ref{fig:sigFeyn} therefore depend on the choice of $x$.

\begin{itemize}
	
	\item $ x = 0.25 $: the mass of the lepton superpartner is
	closer to that of the $\PSGczDo$ than to that of the
	$\PSGcmDo$. Hence, the upper left diagram in
	Fig.~\ref{fig:sigFeyn} produces lower energy tau leptons
	than the upper right. The lower two diagrams both typically
	produce two tau leptons with a large difference in energy.
	
	\item $ x = 0.75 $: the masses of the \PSGtpmDo
	and the \PSGcpmDo are relatively close, so the upper left
	diagram in Fig.~\ref{fig:sigFeyn} produces more energetic
	tau leptons than the upper right. The lower two diagrams
	produce the same energy asymmetry as in the case of $ x=0.25 $.
	
	\item $ x = 0.5 $: the tau leptons in all four diagrams have similar energies.
\end{itemize}

In fact, when all four diagrams are taken into
account the distributions of the kinematic properties are found to be very similar
for the three different values of $ x $, for a given set of chargino and LSP masses.

It is important to note, however, that the choice of chargino mass does
affect the overall sensitivity. For instance, if the chargino is
very close in mass to the top squark, then the
momenta of the \PQb jets are reduced and those of the remaining decay
products are increased. This results in an increase in the
overall sensitivity, provided the \PQb jet \pt values are within the acceptance.
On the other hand if the chargino is very close in mass to the LSP,
then an overall loss of sensitivity is expected. Such scenarios
are not explored in this paper, where the default chargino mass given
in Eq.~(\ref{eq:mass}) is taken throughout.
The polarizations of the tau leptons originating from SUSY cascade decays, which have been found to be useful for studying SUSY signals \cite{Guchait:2002xh}, have not been exploited here.

\section{Event reconstruction}
\label{sec:EventReco}

The particle-flow (PF) algorithm~\cite{CMS-PRF-14-001} aims at
reconstructing each individual particle in an event, with an optimized
combination of information from the various components of the CMS detector. The energy
of photons is obtained from the ECAL measurement, whereas
the momentum of electrons is determined from a combination of the measurement
of momentum by the tracker, the energy of matching ECAL
deposits, and the energy of all bremsstrahlung photons consistent with
originating from the track.
The momentum of muons is obtained from the curvature of the corresponding track.
The energy of charged hadrons is determined from a combination of the
momentum measured in the tracker and the matching ECAL and HCAL
energy deposits, corrected for zero-suppression effects and for the
response function of the calorimeters to hadronic showers. Finally,
the energy of neutral hadrons is obtained from the corresponding
corrected ECAL and HCAL energies.

Reconstruction of jets is performed by clustering PF objects
using the anti-\kt algorithm~\cite{Cacciari:2008gp,Cacciari:2011ma}
with a distance parameter of $R=0.4$.
Jet momentum is determined as the vectorial sum of all particle momenta in
the jet, and is found in simulation to be, on average, within 5--10\%
of the generated momentum over the whole \pt spectrum and detector acceptance.
Additional proton-proton interactions within the same or nearby bunch
crossings (pileup) can contribute spurious tracks and calorimetric energy
deposits, increasing the apparent jet momentum. In order to mitigate
this effect, tracks identified as originating from pileup vertices
are discarded, and an offset is applied to correct for the
remaining contributions~\cite{Khachatryan:2016kdb}. Jets
are calibrated using both simulation
and data studies~\cite{Khachatryan:2016kdb}.
Additional selection criteria are
applied to each jet to remove those potentially dominated by
instrumental effects or reconstruction failures~\cite{CMS-PAS-JME-16-003}.
Jets with $\pt>20\GeV$ and $\abs{\eta}<2.4$ are used in this analysis.

Vertices reconstructed in an event are required to be within 24\unit{cm} of
the center of the detector in the $ z $ direction, and to have a transverse displacement from the beam line
of less than 2\unit{cm}.
The vertex with the largest value of summed physics-object $\pt^2$ is taken to be the primary $\Pp\Pp$ interaction vertex. The physics objects
used for this purpose are jets, clustered using the aforementioned jet finding algorithm with the tracks assigned to the vertex as inputs, and the associated missing transverse momentum, taken as the negative vector sum of the \pt of those jets.

Jets originating from the fragmentation of \PQb quarks are identified as \PQb-tagged jets by using the combined secondary vertex (CSVv2) algorithm~\cite{BTV-16-002},
which utilizes information from displaced tracks and reconstructed secondary vertices. An operating point is chosen corresponding to a signal efficiency of 70\% with a mistagging probability of about 1\% for light jets (from up, down and strange quarks, and gluons) and 15\% for jets originating from charm quarks.

The momentum resolution for electrons with $\pt \approx 45\GeV$
from $\PZ \rightarrow \Pe \Pe$ decays ranges from 1.7 to 4.5\%.
It is generally better in the barrel region than in the endcaps, and
also depends on the bremsstrahlung energy emitted by the electron
as it traverses the material in front of the ECAL~\cite{Khachatryan:2015hwa}.
Electrons with $ \pt > 20\GeV$ and $ \abs{\eta} < 2.4 $ are used for this analysis.

Muons are measured with
detection planes made using three technologies: drift tubes, cathode strip
chambers, and resistive plate chambers.
Matching muons to tracks measured in the silicon tracker results in a \pt
resolution of 1\% in the barrel
and 3\% in the endcaps, for muons with a \pt of up to 100\GeV. The \pt
resolution in the barrel is better than 7\% for muons with a \pt of up to 1\TeV~\cite{Sirunyan:2018_1804.04528}.
This search uses muons with $ \pt > 20 $\GeV and $ \abs{\eta} < 2.4 $.

Isolation criteria are imposed on the lepton (electron and muon)
candidates to reject leptons originating from hadronic decays.
The isolation variable used for this purpose is defined as the scalar
sum of the \pt of reconstructed charged and neutral particles within a
cone of radius $ \Delta R = \sqrt{\smash[b]{(\Delta\eta)^2+(\Delta\phi)^2}}$
 = 0.3\ (0.4) around the electron (muon) candidate track, excluding the
lepton candidate, divided by the \pt of the lepton candidate.
Charged particles not originating from the primary vertex are excluded from this sum and a correction is applied to account for the neutral components originating from pileup, following the procedure described in Ref.~\cite{Khachatryan:2015hwa}.
This relative isolation is required to be less than 15 (20)\% for electron (muons).
The electron and muon candidates passing the aforementioned criteria
are used to identify a control region (CR) that is used for the estimation
of the background from top quark pair production, as explained in Section \ref{sec:ttbarBkg}.

The missing transverse momentum vector \ptvecmiss is computed as the
negative vector sum of the \pt of all the PF candidates
in an event, and its magnitude is denoted as
\ptmiss. The \ptvecmiss is modified to account
for the energy calibration of the reconstructed jets in the event.
The energy calibration of the PF candidates that have not been clustered into jets is also taken into account.
Anomalous high-\ptmiss events may appear because of a variety of reconstruction
failures, detector malfunctions, or backgrounds not originating from
collisions (e.g., particles in the beam halo). Such events
are rejected by filters that are designed to identify more
than 85--90\% of the spurious high-\ptmiss events with a misidentification
rate of less than 0.1\%~\cite{Sirunyan:2019kia}.
In order to minimize the effect of extra noise in the ECAL endcap in
2017, forward jets with uncalibrated $ \pt < 50 $ GeV and $ 2.65 < \abs{\eta} < 3.14 $
are removed from the calculation of \ptmiss in both data and simulation. This improves the agreement between simulation and data at the cost of degrading the \ptmiss resolution by only a few percent.

The hadrons-plus-strips algorithm \cite{Sirunyan:CMS-TAU-16-003}
is used to reconstruct \tauh candidates:
one charged hadron and up to two neutral pions, or three charged
hadrons, consistent with originating from the decay of a tau lepton.
The probability of an electron or muon being misidentified as a \tauh candidate is
greatly reduced by combining
information from the tracker, calorimeters, and muon detector.
The isolation of the \tauh candidate is determined from the
presence of reconstructed particles within a radius
of $\Delta R = 0.3$ around
the \tauh axis that are not
compatible with the decay, and is a useful quantity to distinguish
between jets and \tauh decays.
In order to distinguish between jets originating from quarks or
gluons, and genuine hadronic tau lepton decays, a multivariate
discriminant is calculated from information including the isolation
and measured lifetime.
The \tauh candidates are selected with $ \pt > 40\GeV$, $ \abs{\eta} <
2.1 $, and the ``tight" working point of the above discriminant.
This working point has an efficiency of $ {\approx} $50\% with a
misidentification  probability of $ {\approx} $0.03\%.
The ``loose" working point, which has an efficiency of $ {\approx} $65\% and a misidentification probability of $ {\approx} $0.07\%, is used for estimating the background from misidentified \tauh candidates.

\section{Event selection}
\label{sec:EventSel}

The sources of \ptmiss in the signal events are the neutrinos and the weakly interacting neutralinos, which are correlated with the visible objects (in particular
the $\tauh$ decays).
In contrast, \ptmiss in the SM background processes is primarily due to neutrinos.
This difference can be exploited by first constructing the transverse
mass \mT, defined as follows:
\begin{linenomath}
	\begin{equation}
	\begin{aligned}
	\mT^{2}(\vec{\pt}^\text{vis},\vec{\pt}^\text{inv}) &=
	m^{2}_\text{vis} + m^{2}_\text{inv} + 2(\et^\text{vis} \et^\text{inv}
	- \vec{\pt}^\text{vis} \cdot \vec{\pt}^\text{inv}) , \\
	\text{where} \qquad \et^{2} &= m^{2} + \pt^{2} .
	\end{aligned}
	\label{eq:mT}
	\end{equation}
\end{linenomath}
Here the masses of the visible (vis) and invisible (inv) particles are denoted by $ m_{\text{vis}} $ and $ m_{\text{inv}} $, respectively.
The value of \mT has a maximum at the mass of the parent of the visible and the invisible particles.
To account for multiple sources of missing momentum in the signal
process, the ``stransverse mass" \cite{Lester:1999tx, Barr:2003rg} is defined as:
\begin{linenomath}
	\begin{equation}
	\begin{aligned}
	\mTii^{2} (\text{vis1}, \text{vis2}, \ptmiss) &= \min_{\vec{\pt}^\text{\text{inv1}} + \vec{\pt}^\text{\text{inv2}} = \ptvecmiss} [\max \{ \mT^{2}(\vec{\pt}^\text{\text{vis1}},\vec{\pt}^\text{\text{inv1}}), \mT^{2}(\vec{\pt}^\text{\text{vis2}},\vec{\pt}^\text{\text{inv2}}) \}] .
	\end{aligned}
	\label{eq:mT2}
	\end{equation}
\end{linenomath}
Since the momenta of the individual invisible particles in
Eq.~(\ref{eq:mT2}) are unknown, \ptvecmiss is divided into two
components ($ \vec{\pt}^\text{\text{inv1}} $ and $\vec{\pt}^\text{\text{inv2}} $) in such a way that the value of \mTii is minimized.
If \mTii is computed using the two $\tauh$ candidates as the
visible objects, vis1 and vis2, then its upper limit in the signal will be at the chargino mass.
This is different from the SM background processes. For example in \ttbar events, the upper limit is at the \PW boson mass.
For this analysis, \mTii is calculated with the masses of the invisible
particles in Eq.~(\ref{eq:mT}) set to zero \cite{Barr:2009wu}.

The signal and background processes can be further separated by utilizing the total visible momentum of the system. This is characterized using the quantity \HT, which is defined as the scalar sum of the \pt of all jets and the \tauh candidates in the event. Jets lying within a cone of $ \Delta R = 0.3 $ around either of the two selected $\tauh$ candidates are excluded from this sum to avoid double counting. Being a measure of the total energy of the system, \HT is sensitive to the mass of the top
squark.

\begin{figure}[!htbp]
	\centering
	
	\includegraphics[width=0.495\textwidth]{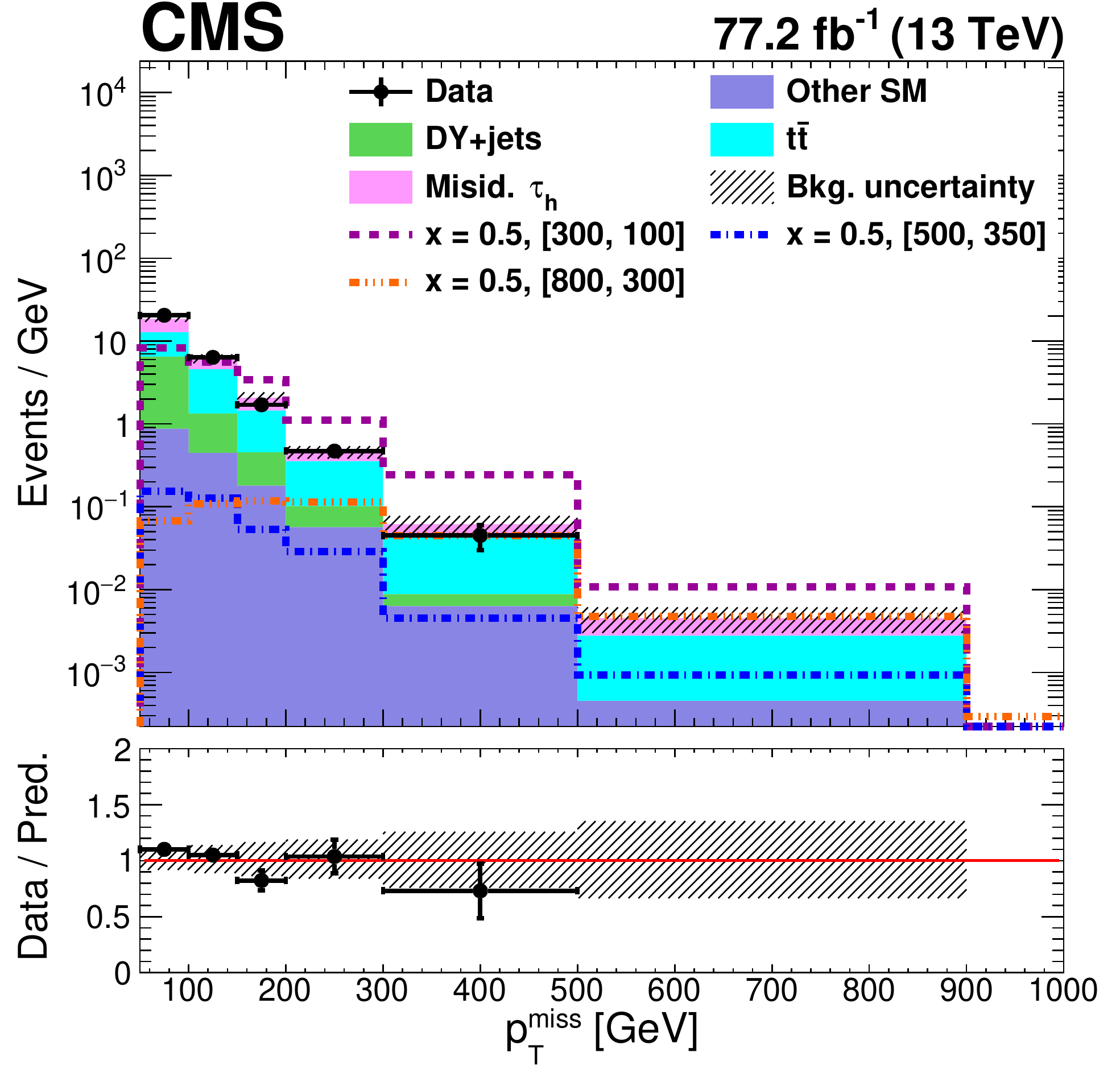}
	\includegraphics[width=0.495\textwidth]{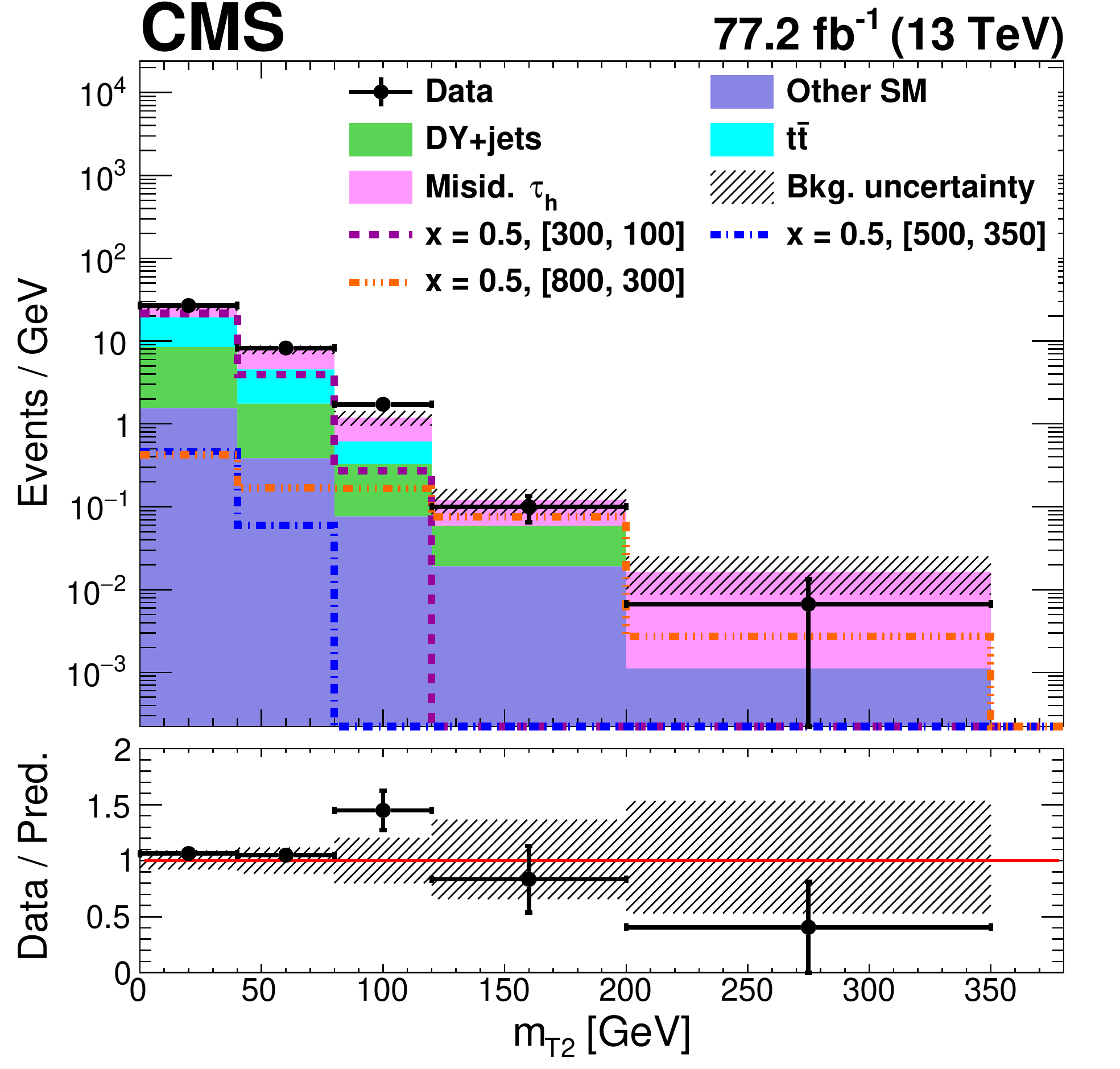} \\
	\includegraphics[width=0.495\textwidth]{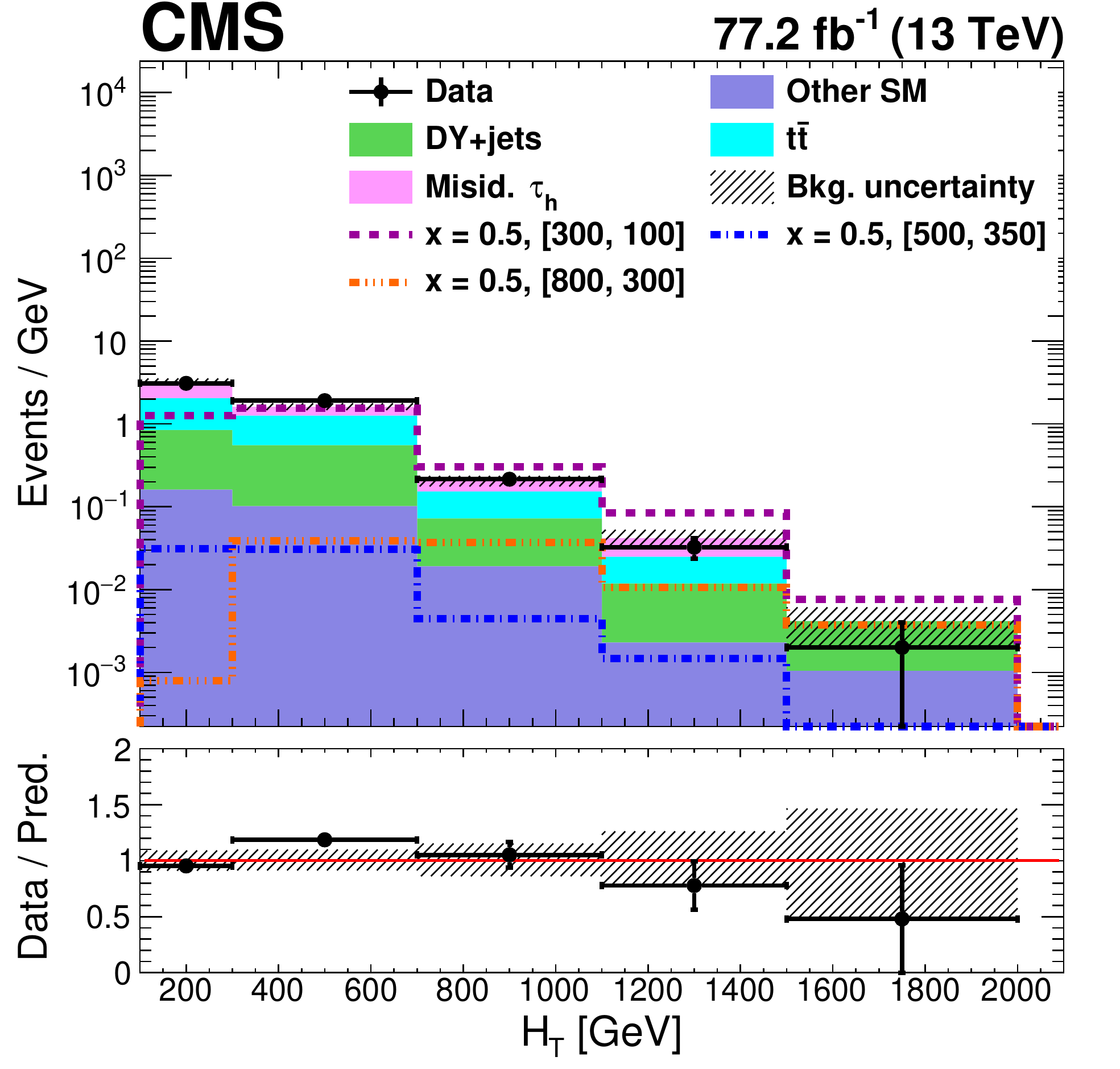}
	
	\caption{
		Distributions of the search variables \ptmiss, \mTii,
		and \HT after event selection, for data and the predicted background.
		The histograms for the background processes are stacked, and the distributions for a few representative signal points corresponding to $ x = 0.5 $ and [$ m_{\PSQtDo} $, $ m_{\PSGczDo} $] = [300, 100], [500, 350], and [800, 300]\GeV are overlaid.
		The lower panel indicates the ratio of the observed data to the background prediction.
		The shaded bands indicate the statistical and systematic uncertainties on the background, added in quadrature.
	}
	\label{fig:SRvariableDataMC201617}
	
\end{figure}

Signal events are selected using \tauhtauh triggers, where both
\tauh candidates are required to have $ \abs{\eta} < 2.1 $, and $ \pt > 35
$ or 40 \GeV depending on the trigger path. The \tauhtauh trigger has an
efficiency of ${\approx} $95\% for \tauh candidates that pass the offline selection.
The trigger efficiencies in simulation are corrected to match the efficiencies measured in data.
For the offline selection, signal events are required to have $\ptmiss
> 50\GeV$, $\HT > 100 \GeV$, at least two oppositely charged \tauh
candidates with $ \pt > 40\GeV$ and $ \abs{\eta} < 2.1 $, and
at least one \PQb-tagged jet with $ \pt > 20\GeV$ and $ \abs{\eta} < 2.4 $.
The requirements on \ptmiss and the number of \PQb-tagged jets ($ n_{\Pb} $) help to
reduce the contributions from DY+jets and SM events comprised uniquely
of jets produced through the strong interaction, referred to as multijet events.
Distributions of the variables \ptmiss, \mTii, and \HT after this
selection are shown in Fig.~\ref{fig:SRvariableDataMC201617} for data and the predicted background,
along with representative signal distributions. The background
prediction
includes \ttbar, DY+jets, events with misidentified \tauh, and other rare SM processes.
Detailed descriptions of the background estimation methods are presented in Section~\ref{sec:BkgEstimation}.

\begin{figure}[!htbp]
	\centering
	
	\includegraphics[width=0.65\textwidth]{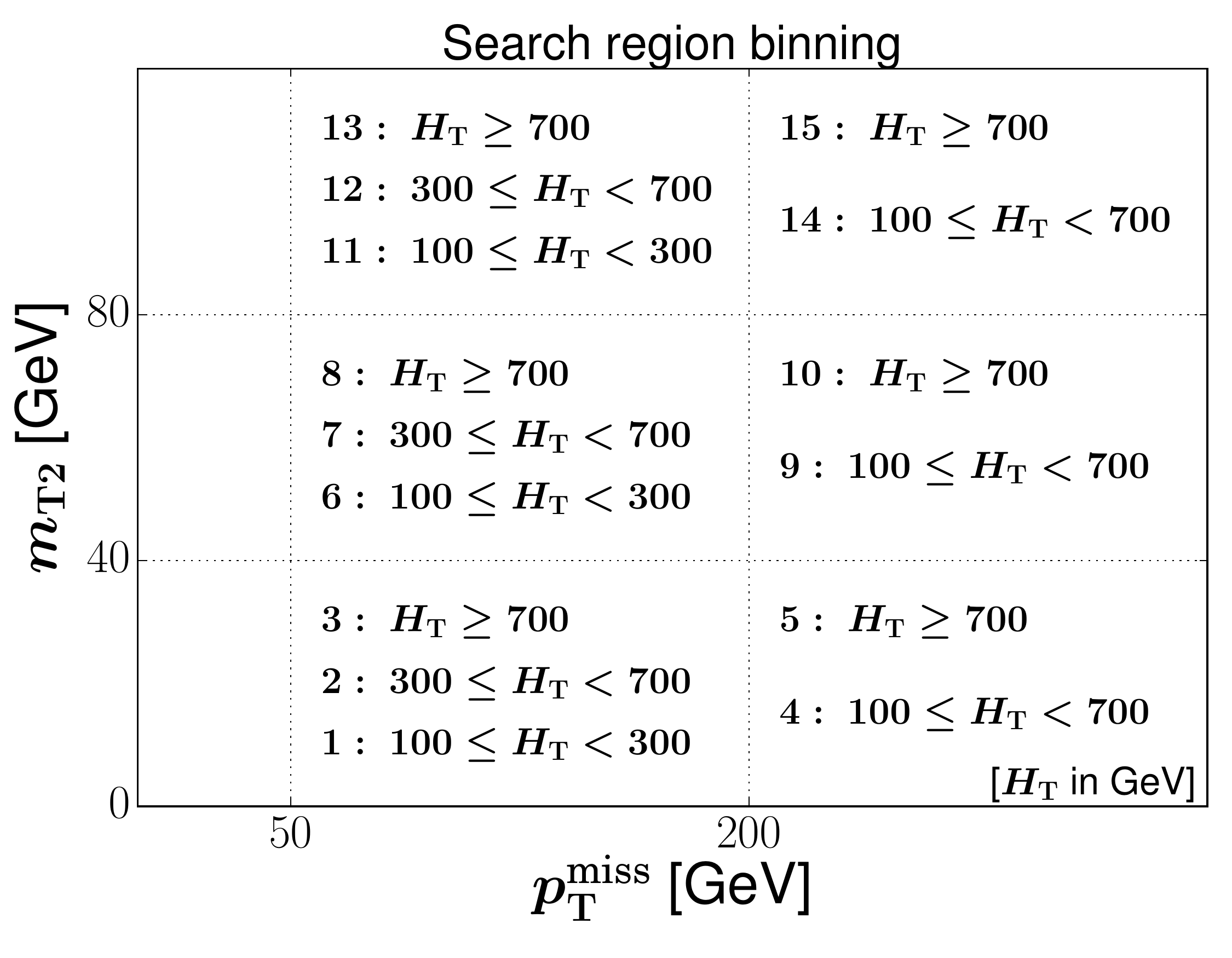}
	
	\caption{The 15 search regions defined in bins of \ptmiss, \mTii, and \HT.}
	\label{fig:sigBin}
\end{figure}

Signal events with different top squark and LSP masses populate different regions of the phase space.
For example, regions with low \ptmiss, \mTii, and \HT are sensitive to signals with
low top squark masses. On the other hand, events with high \ptmiss,
\mTii, and \HT are sensitive to models with high top squark and low
LSP masses. In order to obtain the highest sensitivity over the
entire phase space, the signal region (SR) is divided into 15 bins as
a function of the measured \ptmiss, \mTii, and \HT, which
are illustrated in Fig.~\ref{fig:sigBin}.

\section{Background estimation}
\label{sec:BkgEstimation}

The most significant background is \ttbar production, either with two
genuine \tauh decays or because of jets being misidentified as \tauh candidates.
Because of theoretical uncertainties in the \ttbar background modeling in
the SR (which contains events that populate the tails of the kinematic distributions), we estimate the \ttbar contribution to events with two genuine
\tauh decays using CRs in data, as discussed below.
The background contribution from DY events is typically minor in the most sensitive bins, and has been estimated using simulation.
To account for residual discrepancies between data and the LO DY sample,
correction factors for simulated events are derived from DY-enriched dimuon CRs in data and simulation as functions of the dimuon invariant mass and \pt.
The contribution from multijet events is negligible because of the selections $ \ptmiss > 50\GeV $ and $ n_{\Pb} \geq 1 $.
Other less significant backgrounds, such as {\PW}+jets, $ \PV\PV $, $ \PV\PH $, and
$ \ttbar\PV $ are also estimated from simulation.
The overall SM contribution from jets being misidentified as \tauh
candidates is estimated using CRs in data.
In the following sections we detail the estimation of those
backgrounds that are obtained from CRs in data.

\subsection{Tau lepton pairs from top production}
\label{sec:ttbarBkg}

\begin{figure}[!htbp]
	\centering
	
	\includegraphics[width=0.495\textwidth]{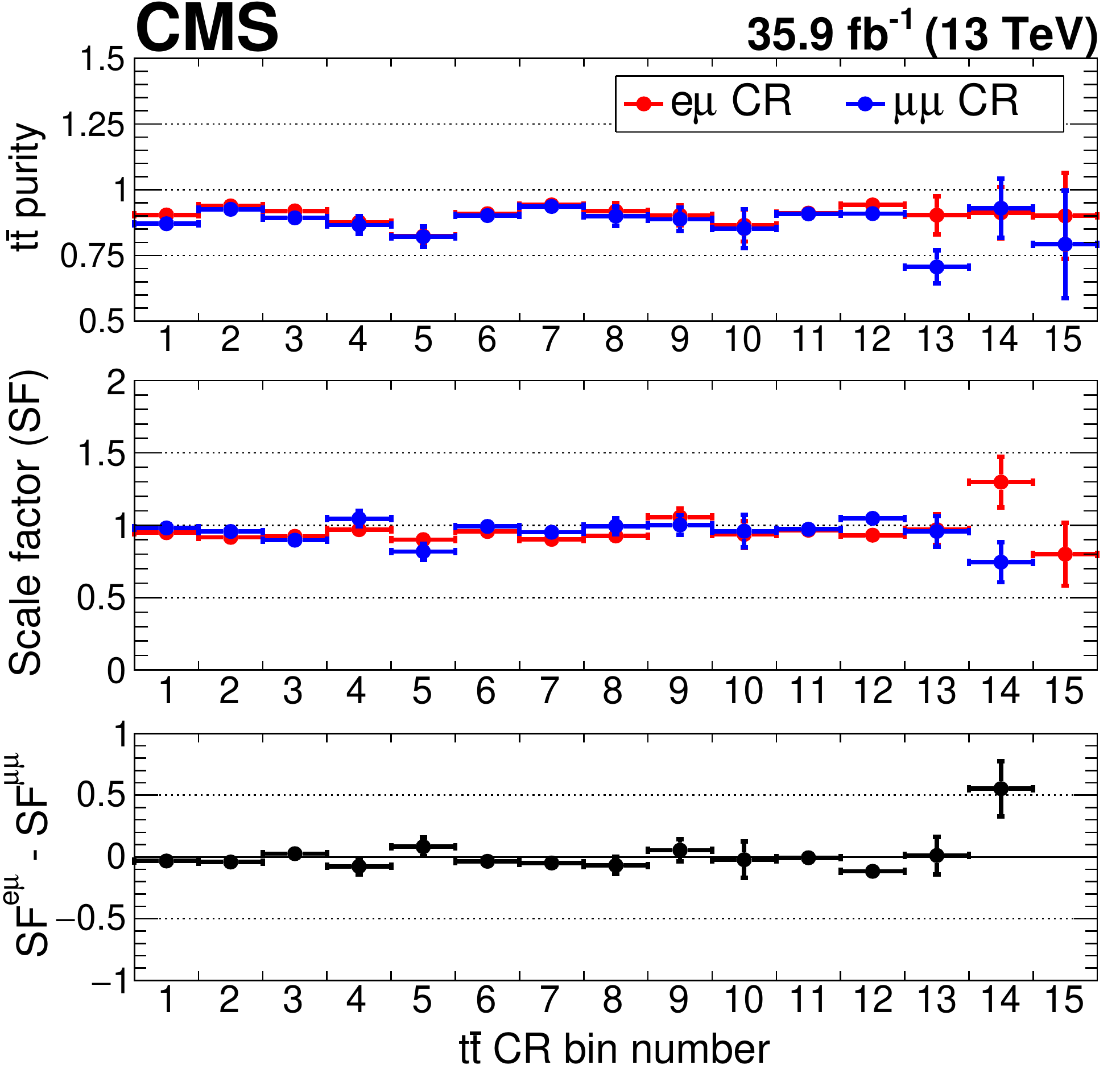}
	\includegraphics[width=0.495\textwidth]{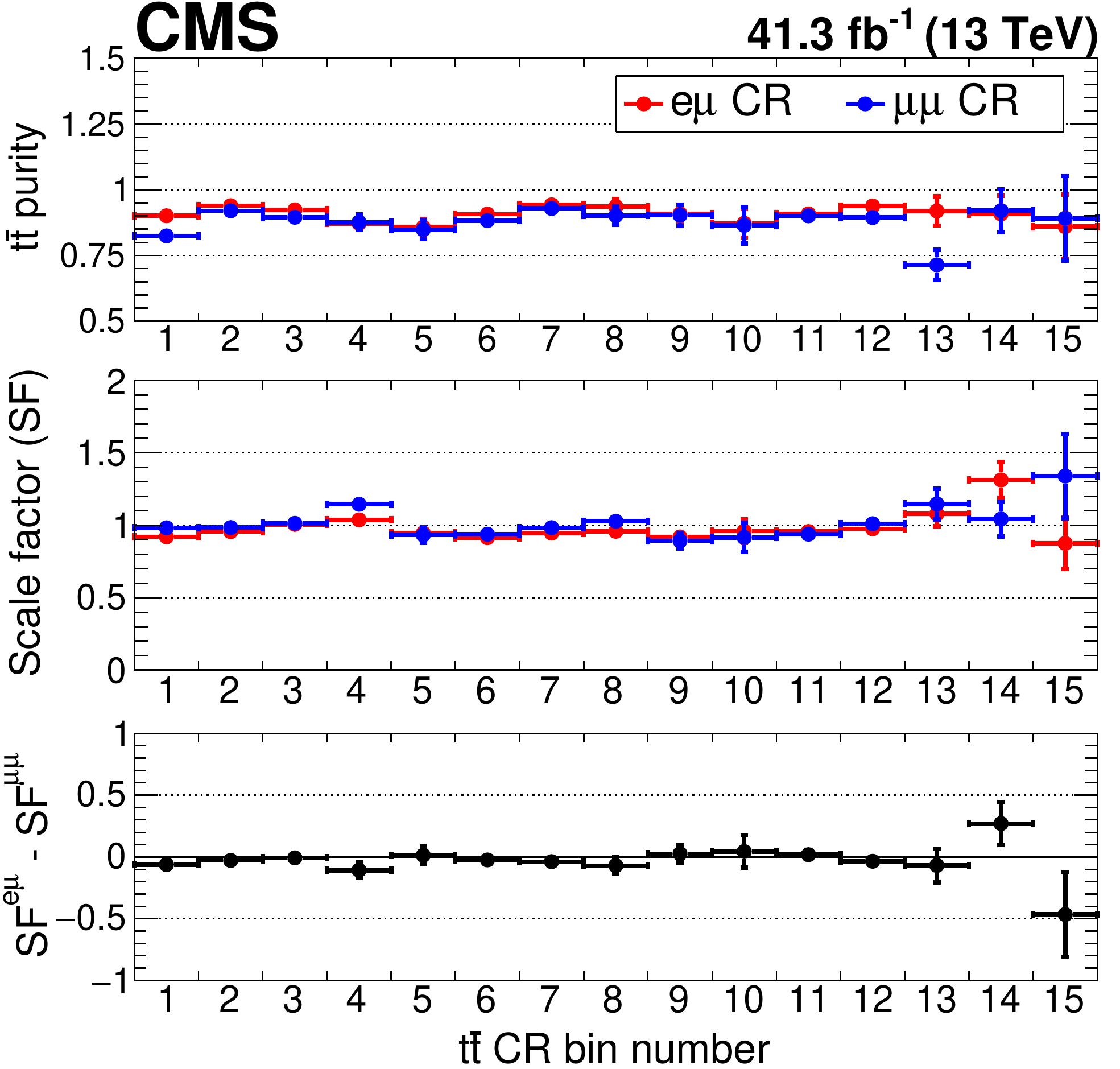}
	
	\caption{The purities (upper row), scale factors
		(middle row), and $ \text{SF}^{\Pe \mu} - \text{SF}^{\mu \mu}
		$ (bottom row) in the different bins (as defined in
		Fig.~\ref{fig:sigBin}) of the \ttbar CR for
		2016 (left) and 2017 (right) data. The scale factor in bin 15 of the
		\mumu CR in 2016 is off the visible scale. Note that bins 14
		and 15 are merged to provide a single SF for subsequent calculations.
		\label{fig:ttbarCRpuritySF}}
\end{figure}

The estimation of the background from \ttbar events in which there
are two genuine \tauh decays
is based on the method described in Ref.~\cite{Sirunyan:2645851}.
The predicted yields in each SR bin from simulation are multiplied by
correction factors derived in a {\ttbar}-enriched CR.
The \ttbar-enriched CR is identified by selecting events with an oppositely charged \emu pair.
These events are selected with \emu triggers, and are required to satisfy the same
offline requirements as the SR with the \Pe and \Pgm
replacing the two \tauh candidates. The \emu triggers are $ {\approx}$95\% efficient for lepton candidates.
In addition, in order to reduce possible DY contamination (from the
tail of the \emu invariant mass distribution in the process $ \PZ/\gamma^{*} \to \tau\tau \to \emu $) in this CR, events are vetoed if the
invariant mass of the \emu system lies in the range $ 60 < m_{\emu} <
120\GeV$. This selection on the dilepton invariant mass is more effective in the \mumu CR to be discussed later, but is also applied here in order to be consistent.
Other objects, such as jets and \PQb-tagged jets, are selected using the
same kinematic requirements and working points as in the SR.
The definitions of the search variables for this CR are the same as those in the SR, except that the \emu pair is used in place of the \tauh pair for evaluating the search variables.
The purity of \ttbar in the CR
(i.e., the fraction of \ttbar events in each bin) is measured in
simulation as $ {\gtrsim} $85\% , as shown in
Fig.~\ref{fig:ttbarCRpuritySF} (upper panels).

Residual differences between
data and simulation are quantified by SFs.
For a given SR region ($ i $) we define
\begin{linenomath}
	\begin{equation}
	\begin{aligned}
	SF_{i} &= \frac{N^{\emu \ \text{CR}}_{i,\ \text{data}}}{N^{\emu \ \text{CR}}_{i,\ \text{MC}}} ,
	\end{aligned}
	\label{eq:ttbarSF}
	\end{equation}
\end{linenomath}
where the numerator and the denominator represent the yields in the CR
in data and simulation, respectively.
The corrected \ttbar yield in simulation in each region of the SR is then obtained as:
\begin{linenomath}
	\begin{equation}
	\begin{aligned}
	N^{\tauhtauh \ \text{SR}}_{i,\ \text{corr } \ttbar} &=
	N^{\tauhtauh \ \text{SR}}_{i,\ \ttbar \ \text{MC}} SF_{i} &=
	\frac{N^{\emu \ \text{CR}}_{i,\ \text{data}} \ N^{\tauhtauh \ \text{SR}}_{i,\ \ttbar \ \text{MC}}}{N^{\emu \ \text{CR}}_{i,\ \text{MC}}}
	,
	\end{aligned}
	\label{eq:ttbarSFcorr}
	\end{equation}
\end{linenomath}
where $ N^{\tauhtauh \ \text{SR}}_{i,\ \ttbar \ \text{MC}} $ is the
prediction from simulated \ttbar events in the SR.
An alternative way of interpreting this method is that we take
the \ttbar spectrum from a \ttbar-enriched \emu CR in data ($ N^{\emu
	\ \text{CR}}_{i,\ \text{data}} $) and extrapolate it to the $
\tauhtauh $ SR by accounting for the differences between the
properties of $ \tauhtauh $ and \emu final states with the ratio $
N^{\tauhtauh \ \text{SR}}_{i,\ \ttbar \ \text{MC}} / N^{\emu \
	\text{CR}}_{i,\ \text{MC}} $ taken from simulation.
The SFs in the different bins, shown in
Fig.~\ref{fig:ttbarCRpuritySF} (middle row) for both 2016 and 2017 data,
are mostly found to be within $ {\approx} $10\% of unity. Note that
separate SFs for bins 14 and 15 are shown for information, but
these are merged and a single SF is used in subsequent calculations
to reduce the statistical uncertainty.

In order to cross-check the validity of this method, the same technique is applied to an independent \ttbar-enriched CR
with an oppositely charged \mumu pair in the final state. These events are selected with single
muon triggers that reach $ {\approx} $95\% efficiency.
The event selection for the \mumu CR is similar to that for the $\emu$ CR.
This cross-check evaluates the effect of possible contamination from
DY events (the branching fraction of $ \PZ/\gamma^{*} \to \Pgm\Pgm $ being much higher than that of $ \PZ/\gamma^{*} \to \tau\tau \to \emu $),
and is also useful for checking any dependence of the SFs on lepton reconstruction.
The differences between the SFs calculated in the main and cross-check
CRs, shown in Fig.~\ref{fig:ttbarCRpuritySF} (bottom row), are small (within
$ {\approx}$10\% in most cases), and are taken as an uncertainty in the
SFs. These are added in quadrature to the statistical uncertainty in
the SFs, and propagated as a contribution to the uncertainty in the final \ttbar prediction.

\subsection{Misidentified hadronically decaying tau lepton candidates}

The next largest component of the total background originates from quark or gluon jets that are misidentified as a \tauh candidate.
The largest sources of such events in the SR are semileptonic and hadronic \ttbar decays. We estimate this contribution to the SR following a strategy~\cite{Sirunyan:2628769} that uses the yields in {\tauh}\tauh CRs, defined by inverting the requirements on the working point of the \tauh identification.

For a genuine \tauh passing the loose identification requirements, we
define $g$ as the probability that it also passes the tight identification requirements.
We define $f$ as the corresponding probability for a misidentified
\tauh candidate.
We then define $ N_\text{gf}$ as the number of {\tauh}\tauh events
where the \tauh candidate with the highest \pt is genuine and that
with the second-highest \pt is misidentified, with other terms ($ N_\text{fg}$, $ N_\text{gg}$, and $ N_\text{ff}$) defined similarly. We also define
$N_\text{TL}$ as the number of {\tauh}\tauh events where the candidate
with the highest \pt passes the tight identification criteria and
that with the second-highest \pt fails, but
passes the loose criteria, with other terms ($N_\text{LT}$, $N_\text{LL}$, and $N_\text{TT}$) defined similarly.
If $ N $ is the total number of events, the following set of equations can be constructed:
{\small
	\begin{linenomath}
		\begin{equation}
		\begin{aligned}
		N &= N_\text{gg} + N_\text{fg} + N_\text{gf} + N_\text{ff} = N_\text{TT} + N_\text{LT} + N_\text{TL} + N_\text{LL}, \\
		N_\text{LL} &= (1-g_{1})(1-g_{2}) N_\text{gg} + (1-f_{1})(1-g_{2}) N_\text{fg} + (1-g_{1})(1-f_{2}) N_\text{gf} + (1-f_{1})(1-f_{2}) N_\text{ff}, \\
		N_\text{LT} &= (1-g_{1}) g_{2} N_\text{gg} + (1-f_{1}) g_{2} N_\text{fg} + (1-g_{1}) f_{2} N_\text{gf} + (1-f_{1}) f_{2} N_\text{ff}, \\
		N_\text{TL} &= g_{1} (1-g_{2}) N_\text{gg} + f_{1} (1-g_{2}) N_\text{fg} + g_{1} (1-f_{2}) N_\text{gf} + f_{1} (1-f_{2}) N_\text{ff}, \\
		N_\text{TT} &= g_{1} g_{2} N_\text{gg} + f_{1} g_{2} N_\text{fg} + g_{1} f_{2} N_\text{gf} + f_{1} f_{2} N_\text{ff} ,
		\end{aligned}
		\label{FakeEquation_1}
		\end{equation}
	\end{linenomath}
}%
where the subscripts 1 and 2 on $g$ and $f$ refer to the \tauh
candidates with the highest and
second-highest \pt, respectively.

The above equations can be inverted to give the numbers of genuine and
misidentified {\tauh}\tauh candidate events in the SR:
\begin{linenomath}
	\begin{equation}
	N_\text{TT} = N^{\text{gen}}_\text{TT} + N^{\text{misid}}_\text{TT},
	\end{equation}
	\text{where}
	\begin{equation*}
	\begin{aligned}
	N^{\text{gen}}_\text{TT} &= g_{1} g_{2} N_\text{gg}, \\
	N^{\text{misid}}_\text{TT} &= f_{1} g_{2} N_\text{fg} + g_{1} f_{2} N_\text{gf} + f_{1} f_{2} N_\text{ff} .
	\end{aligned}
	\label{FakeEquation_3}
	\end{equation*}
\end{linenomath}
Here $ N^{\text{gen}}_\text{TT} $ represents the number of events
in the SR with two genuine \tauh candidates in the final state,
and $ N^{\text{misid}}_\text{TT} $ the number of events in the SR with one or two misidentified \tauh candidates.

The probability $g$ is determined using \ttbar simulation, with the \tauh candidate being matched to a generated hadronically decaying tau
within a cone of radius $ \Delta R = 0.3 $.
The value of $ g $ is calculated as the ratio between the number of
genuine \tauh jets passing the tight identification criteria and the number passing the loose criteria.
It is evaluated as a function of the \tauh decay modes and \pt and is observed to be about 80\% with very little dependence on the \pt of the \tauh. The dependence on the decay mode is observed to be at the 10\% level.

The misidentification rate $ f $ is estimated using a multijet-enriched CR in data. This CR is defined by requiring a same-charge \tauh pair  satisfying the \tauh  selection criteria, and by requiring $ \ptmiss < 50\GeV$.
The misidentification rate for a single \tauh candidate is estimated from this CR using the following two definitions:
\begin{linenomath}
	\begin{equation}
	\begin{aligned}
	f(\text{LL} \rightarrow \text{TL}) &= \frac{\tauh^{1}(\text{T}) \ \tauh^{2}(\text{L})}{\tauh^{1}(\text{L}) \ \tauh^{2}(\text{L})} ,\\
	f(\text{TL} \rightarrow \text{TT}) &= \frac{\tauh^{1}(\text{T}) \ \tauh^{2}(\text{T})}{\tauh^{1}(\text{T}) \ \tauh^{2}(\text{L})} .
	\end{aligned}
	\label{eq:fakeRate}
	\end{equation}
\end{linenomath}
Here, the term $ \tauh^{i}(\text{X})$ denotes the number of events
where the candidate with the highest ($i=1$) or
second-highest ($i=2$) \pt
passes the tight ($ \text{X=T} $) or loose ($\text{X=L}$) identification criteria.
In each of the two definitions above, the working point of one of the
\tauh candidates in the numerator is changed with respect to the denominator,
so they could be expected to yield the same result. However, if the
probability of one \tauh candidate passing the tight criteria is correlated with
the probability of the other to pass, differences may occur.
In practice, differences of up to $ {\approx}$10\%, depending on the \pt and the decay mode of the \tauh, are observed between the two definitions.
These differences are used to estimate the uncertainty in this method.

The misidentification rate is measured as a function of the \tauh decay modes and \pt. It is found to be around 35\% with a mild dependence on the \pt of the \tauh candidate.
The variations with decay mode are up to the  20\% level.
It has been found in simulation studies~\cite{Sirunyan:2628769} that
the misidentification rate also depends on the
flavor of the parton corresponding to the jet that is misidentified as a \tauh.
Since the jet flavor cannot be reliably determined in data, an additional 15\% uncertainty in $ f $ is included.
This uncertainty is evaluated as the relative difference between the average and the maximum (or minimum) of the misidentification rates corresponding to the different jet flavors (up, down, strange, and bottom quarks, and gluons), estimated using simulated {\PW}+jets events.

\section{Systematic uncertainties}
\label{sec:Systematics}

There are several sources of systematic uncertainty that are propagated to the prediction of the final signal and background yields.
The most significant is the uncertainty in the modeling of the identification and isolation
requirements (ID-iso)~\cite{Sirunyan:CMS-TAU-16-003} of the \tauh candidates,
estimated to be approximately 10\% for all processes in 2016, and 20\% in 2017.
The other sources of uncertainty affecting all processes include the jet energy scale (JES) and jet energy resolution (JER), the \tauh energy scale, the effect of unclustered components in calculating \ptmiss, pileup reweighting, and the \PQb tagging efficiency. The simulation is reweighted to make its pileup distribution identical to that in data. The pileup in data depends on the measured total inelastic cross section~\cite{Sirunyan:2018nqx}, which is varied by $\pm$2.5\% to obtain the uncertainty in this correction.

Since the \ttbar contribution in the SR is obtained by multiplying the simulated yield by a SF, defined as the ratio between the number of events in data and simulation,
several uncertainties cancel to first order.
As mentioned earlier, the difference between the \ttbar SFs obtained in the \emu and \mumu CRs, added in quadrature with the statistical uncertainty, is taken as the uncertainty in this method.
The difference between the two definitions of the misidentification rate, as defined in Eq.~(\ref{eq:fakeRate}), is taken to be the uncertainty in the misidentification rate, while the flavor dependence of the rate is accounted for by adding an additional 15\% uncertainty.

The factorization ($ \mu_{\text{F}} $) and renormalization ($ \mu_{\text{R}} $) scales used in the simulation are varied up and down by a factor of two, avoiding the cases in which one is doubled and the other is halved. The \SYSCALC package~\cite{Kalogeropoulos:2018cke} has been used for this purpose. The resulting uncertainty is estimated to be less than 6\% for both signal and background processes estimated from simulation.
A 2.5\% uncertainty in the measured integrated luminosity is used for 2016~\cite{CMS-PAS-LUM-17-001}, reducing to 2.3\% for 2017~\cite{CMS-PAS-LUM-17-004}.
The uncertainty in the \PZ boson \pt correction applied to DY+jets events is taken to be equal to the deviation of the correction factor from unity.
A normalization uncertainty of 15\% is assigned to the production cross sections of the background processes that are evaluated directly from simulation \cite{Sirunyan:2018owv, Sirunyan:2018ucr, Aaboud:2017qkn, CMS:2019too, Sirunyan:2019hqb, Sirunyan:2019bez, Sirunyan:2017wgx}.

Since the simulation of the detector for signal events is performed using \textsc{FastSim}, the signal yields
are corrected to account for the differences in the \tauh identification efficiency with respect to the
\GEANTfour simulation used for the backgrounds. The statistical uncertainty in this correction is propagated as its uncertainty.
The \textsc{FastSim} package has a worse \ptmiss resolution than the full \GEANTfour simulation, resulting in a potential artificial enhancement of the signal yields. To account for this, the signal yields are corrected, and the uncertainty in the resulting correction to the yield is estimated to be 5--10\% .

The uncertainties in the signal and background from all sources are presented in
Table \ref{tab:uncertainty20162017}.
Upper and lower numbers correspond to the relative uncertainties due to the upward and downward variations of the respective source.
These values are the weighted averages of the relative uncertainties in the different search bins with the weights being the yields in the respective bins.
The tabulated sources of systematic uncertainties are modeled by log-normal distributions~\cite{CMS-NOTE-2011-005} in the likelihood function used for the statistical interpretation of the results, which is discussed in Section \ref{sec:Results}. These uncertainties are considered not to be correlated with each other, but correlated across the 15 search bins.
In addition, the statistical uncertainties are also taken into account and are considered to be uncorrelated across the bins.

\begin{table}[!htbp]
	\centering
	
	\topcaption{Relative systematic uncertainties from different sources in signal and background yields in the 2016 and 2017 analyses combined. These values are the weighted (by the yields in the respective bins) averages of the relative uncertainties in the different search regions. For the asymmetric uncertainties, the upper (lower) entry is the uncertainty due to the upward (downward) variation, which can be in the same direction as a result of taking the weighted average. The numbers in square brackets in the heading indicate the top squark and LSP masses in \GeV, respectively.}
	
	\cmsTable{
		\begin{tabular}{ l  c  c  c  c  c  c  c }
			\hline
			Uncertainty source &
			\vtop{\hbox{\strut $ x = 0.5 $}\hbox{\strut $ [300, 100] $}} &
			\vtop{\hbox{\strut $ x = 0.5 $}\hbox{\strut $ [500, 350] $}} &
			\vtop{\hbox{\strut $ x = 0.5 $}\hbox{\strut $ [800, 300] $}} &
			$ \ttbar $ &
			DY+jets &
			Other SM &
			Misid. \tauh \\
			
			\hline
			\\[\cmsTabSkip]
			Signal cross section &
			$ \pm 6.7 \% $ &
			$ \pm 7.5 \% $ &
			$ \pm 9.5 \% $ &
			\NA &
			\NA &
			\NA &
			\NA \\[\cmsTabSkip]
			
			\textsc{FastSim} $ \ptmiss $ resolution &
			$ \pm 7.4 \% $ &
			$ \pm 10 \% $ &
			$ \pm 5.1 \% $ &
			\NA &
			\NA &
			\NA &
			\NA \\[\cmsTabSkip]
			
			$ \tauh $ \textsc{FastSim}/{\GEANTfour} &
			\vtop{\hbox{\strut $ +4.4 \% $}\hbox{\strut $ -4.3 \% $}} &
			\vtop{\hbox{\strut $ +3.7 \% $}\hbox{\strut $ -3.6 \% $}} &
			\vtop{\hbox{\strut $ +7.9 \% $}\hbox{\strut $ -7.5 \% $}} &
			\NA &
			\NA &
			\NA &
			\NA \\[\cmsTabSkipLarge]
			
			JER &
			\vtop{\hbox{\strut $ -0.27 \% $}\hbox{\strut $ < $0.1\%}} &
			\vtop{\hbox{\strut $ -0.81 \% $}\hbox{\strut $ < $0.1\%}} &
			\vtop{\hbox{\strut $ < $0.1\%}\hbox{\strut $ < $0.1\%}} &
			\NA &
			\vtop{\hbox{\strut $ +0.47 \% $}\hbox{\strut $ +0.27 \% $}} &
			\vtop{\hbox{\strut $ -0.95 \% $}\hbox{\strut $ -0.29 \% $}} &
			\NA \\[\cmsTabSkipLarge]
			
			JES &
			\vtop{\hbox{\strut $ +0.18 \% $}\hbox{\strut $ -0.57 \% $}} &
			\vtop{\hbox{\strut $ < $0.1\%}\hbox{\strut $ -0.81 \% $}} &
			\vtop{\hbox{\strut $ +0.1 \% $}\hbox{\strut $ < $0.1\%}} &
			\NA &
			\vtop{\hbox{\strut $ +1.1 \% $}\hbox{\strut $ -1.7 \% $}} &
			\vtop{\hbox{\strut $ +0.2 \% $}\hbox{\strut $ -1.6 \% $}} &
			\NA \\[\cmsTabSkipLarge]
			
			$ \mu_{\text{R}} $ and $ \mu_{\text{F}} $ scales &
			\vtop{\hbox{\strut $ +0.6 \% $}\hbox{\strut $ -0.7 \% $}} &
			\vtop{\hbox{\strut $ +1.9 \% $}\hbox{\strut $ -2.1 \% $}} &
			\vtop{\hbox{\strut $ +0.31 \% $}\hbox{\strut $ -0.35 \% $}} &
			\NA &
			\vtop{\hbox{\strut $ +4.6 \% $}\hbox{\strut $ -4.4 \% $}} &
			\vtop{\hbox{\strut $ +3.2 \% $}\hbox{\strut $ -2.6 \% $}} &
			\NA \\[\cmsTabSkipLarge]
			
			$ \tauh $ ID-iso &
			\vtop{\hbox{\strut $ +16 \% $}\hbox{\strut $ -14 \% $}} &
			\vtop{\hbox{\strut $ +16 \% $}\hbox{\strut $ -14 \% $}} &
			\vtop{\hbox{\strut $ +17 \% $}\hbox{\strut $ -15 \% $}} &
			\vtop{\hbox{\strut $ +16 \% $}\hbox{\strut $ -15 \% $}} &
			\vtop{\hbox{\strut $ +16 \% $}\hbox{\strut $ -14 \% $}} &
			\vtop{\hbox{\strut $ +16 \% $}\hbox{\strut $ -14 \% $}} &
			\NA \\[\cmsTabSkipLarge]
			
			Pileup &
			\vtop{\hbox{\strut $ < $0.1\%}\hbox{\strut $ < $0.1\%}} &
			\vtop{\hbox{\strut $ +0.25 \% $}\hbox{\strut $ -0.18 \% $}} &
			\vtop{\hbox{\strut $ -0.69 \% $}\hbox{\strut $ +0.68 \% $}} &
			\NA &
			\vtop{\hbox{\strut $ -0.23 \% $}\hbox{\strut $ +0.24 \% $}} &
			\vtop{\hbox{\strut $ -0.88 \% $}\hbox{\strut $ +0.88 \% $}} &
			\NA \\[\cmsTabSkipLarge]
			
			\ptmiss unclustered energy &
			\vtop{\hbox{\strut $ -0.16 \% $}\hbox{\strut $ -0.78 \% $}} &
			\vtop{\hbox{\strut $ +1 \% $}\hbox{\strut $ -2 \% $}} &
			\vtop{\hbox{\strut $ < $0.1\%}\hbox{\strut $ < $0.1\%}} &
			\NA &
			\vtop{\hbox{\strut $ +7 \% $}\hbox{\strut $ -2 \% $}} &
			\vtop{\hbox{\strut $ +1.3 \% $}\hbox{\strut $ -1.6 \% $}} &
			\NA \\[\cmsTabSkipLarge]
			
			Background normalization &
			\NA &
			\NA &
			\NA &
			\NA &
			$ \pm 15 \% $ &
			$ \pm 15 \% $ &
			\NA \\[\cmsTabSkip]
			
			Luminosity &
			$ \pm 2.4 \% $ &
			$ \pm 2.4 \% $ &
			$ \pm 2.4 \% $ &
			\NA &
			$ \pm 2.4 \% $ &
			$ \pm 2.4 \% $ &
			\NA \\[\cmsTabSkip]
			
			\PQb tagging &
			\vtop{\hbox{\strut $ +1.1 \% $}\hbox{\strut $ -1.2 \% $}} &
			\vtop{\hbox{\strut $ +0.65 \% $}\hbox{\strut $ -0.67 \% $}} &
			\vtop{\hbox{\strut $ +0.73 \% $}\hbox{\strut $ -0.75 \% $}} &
			\NA &
			\vtop{\hbox{\strut $ +4.7 \% $}\hbox{\strut $ -4.7 \% $}} &
			\vtop{\hbox{\strut $ +2.4 \% $}\hbox{\strut $ -2.4 \% $}} &
			\NA \\[\cmsTabSkipLarge]
			
			$ \tauh $ energy scale &
			\vtop{\hbox{\strut $ +2 \% $}\hbox{\strut $ -3 \% $}} &
			\vtop{\hbox{\strut $ +2.3 \% $}\hbox{\strut $ -3.3 \% $}} &
			\vtop{\hbox{\strut $ +1.1 \% $}\hbox{\strut $ -0.9 \% $}} &
			\vtop{\hbox{\strut $ +1.5 \% $}\hbox{\strut $ -1.8 \% $}} &
			\vtop{\hbox{\strut $ +1.2 \% $}\hbox{\strut $ -1.6 \% $}} &
			\vtop{\hbox{\strut $ +0.7 \% $}\hbox{\strut $ -1.3 \% $}} &
			\NA \\[\cmsTabSkipLarge]
			
			$ \ttbar $ SF &
			\NA &
			\NA &
			\NA &
			$ \pm 2.5 \% $ &
			\NA &
			\NA &
			\NA \\[\cmsTabSkip]
			
			\PZ $ \pt $ reweighting &
			\NA &
			\NA &
			\NA &
			\NA &
			\vtop{\hbox{\strut $ +9.1 \% $}\hbox{\strut $ -9.1 \% $}} &
			\NA &
			\NA \\[\cmsTabSkipLarge]
			
			$ \tauh $ misid. rate (parton flavour) &
			\NA &
			\NA &
			\NA &
			\NA &
			\NA &
			\NA &
			\vtop{\hbox{\strut $ +16 \% $}\hbox{\strut $ -17 \% $}} \\[\cmsTabSkipLarge]
			
			\vtop{\hbox{\strut $ \tauh $ misid. rate}\hbox{\strut ($ \text{LL} \to \text{TL} $ vs. $ \text{TL} \to \text{TT} $)}} &
			\NA &
			\NA &
			\NA &
			\NA &
			\NA &
			\NA &
			\vtop{\hbox{\strut $ +2.7 \% $}\hbox{\strut $ -2.5 \% $}} \\[\cmsTabSkipLarge]
			
			\hline
		\end{tabular}
	}
	
	\label{tab:uncertainty20162017}
\end{table}

\section{Results}
\label{sec:Results}

\begin{figure}[!htbp]
	\centering
	
	\includegraphics[width=0.75\textwidth]{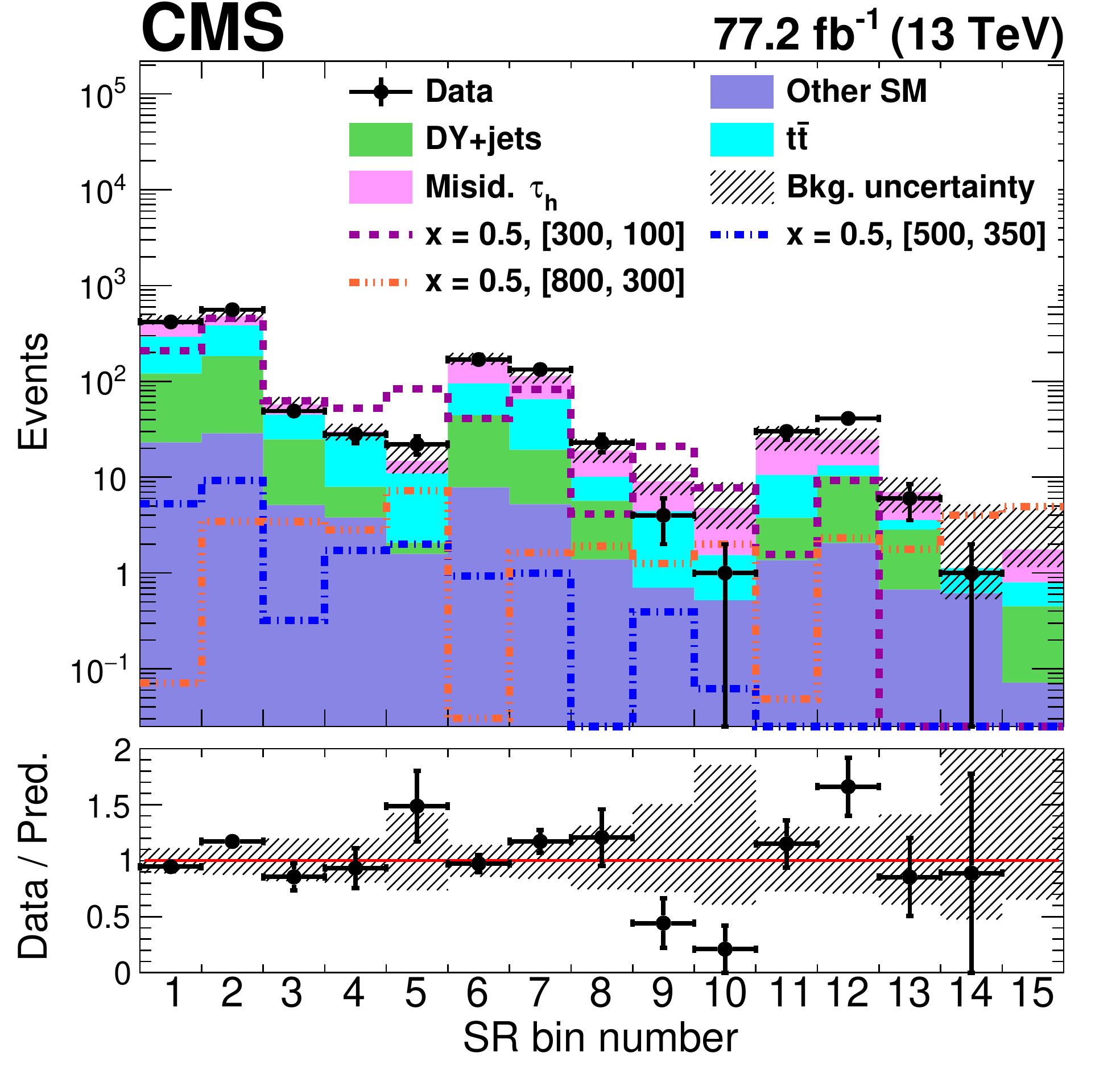}
	\caption{
		Event yields in the 15 search bins as defined in
		Fig.~\ref{fig:sigBin}. The yields for the background processes are stacked, and those for a few
		representative signal points corresponding to $ x =
		0.5 $ and $[ m_{\PSQtDo},  m_{\PSGczDo} ] = [300,
		100]$, [500, 350], and [800, 300]\GeV are overlaid. The
		lower panel indicates the ratio of the observed data
		to the background prediction in each bin.
		The shaded bands indicate the statistical and systematic uncertainties in the background, added in quadrature.
	}
	\label{fig:BGyield}
	
\end{figure}

We present the observed and expected yields in all 15 search bins in
Table \ref{tab:Bg20162017} along with their uncertainties.
Figure \ref{fig:BGyield} shows the observed data in all of the search
bins, compared to the signal and background predictions.

As expected, the dominant contributions in the sensitive
signal bins are from \ttbar and misidentified  \tauh backgrounds.
In cases where the background prediction of a process in a given bin is negligible, the statistical uncertainty is modeled by a gamma distribution~\cite{CMS-NOTE-2011-005} in the likelihood function used for the statistical interpretation, and the Poissonian upper limit at 68\% confidence level (\CL) is shown as a positive uncertainty in the table.
The number of events observed in data is found to be consistent with
the SM background prediction.

The test statistic used for the interpretation of the result is the profile likelihood ratio $ q_{\mu} = -2 \ln{(\mathcal{L}_{\mu} / \mathcal{L}_{\text{max}})} $, where $ \mathcal{L}_{\mu} $ is the maximum likelihood for a fixed signal strength $ \mu $, and $ \mathcal{L}_{\text{max}} $ is the global maximum of the likelihood~\cite{CMS-NOTE-2011-005}.
We set upper limits on signal production at 95\% \CL using a modified frequentist approach and the \CLs
criterion~\cite{JUNK1999435, Read_2002}, implemented through an asymptotic approximation of the test statistic \cite{Cowan2011}.
In this calculation all the background and signal uncertainties are modeled as nuisance parameters and profiled in the maximum likelihood fit.

\begin{figure}[!htbp]
	\centering
	
	\includegraphics[width=0.495\textwidth]{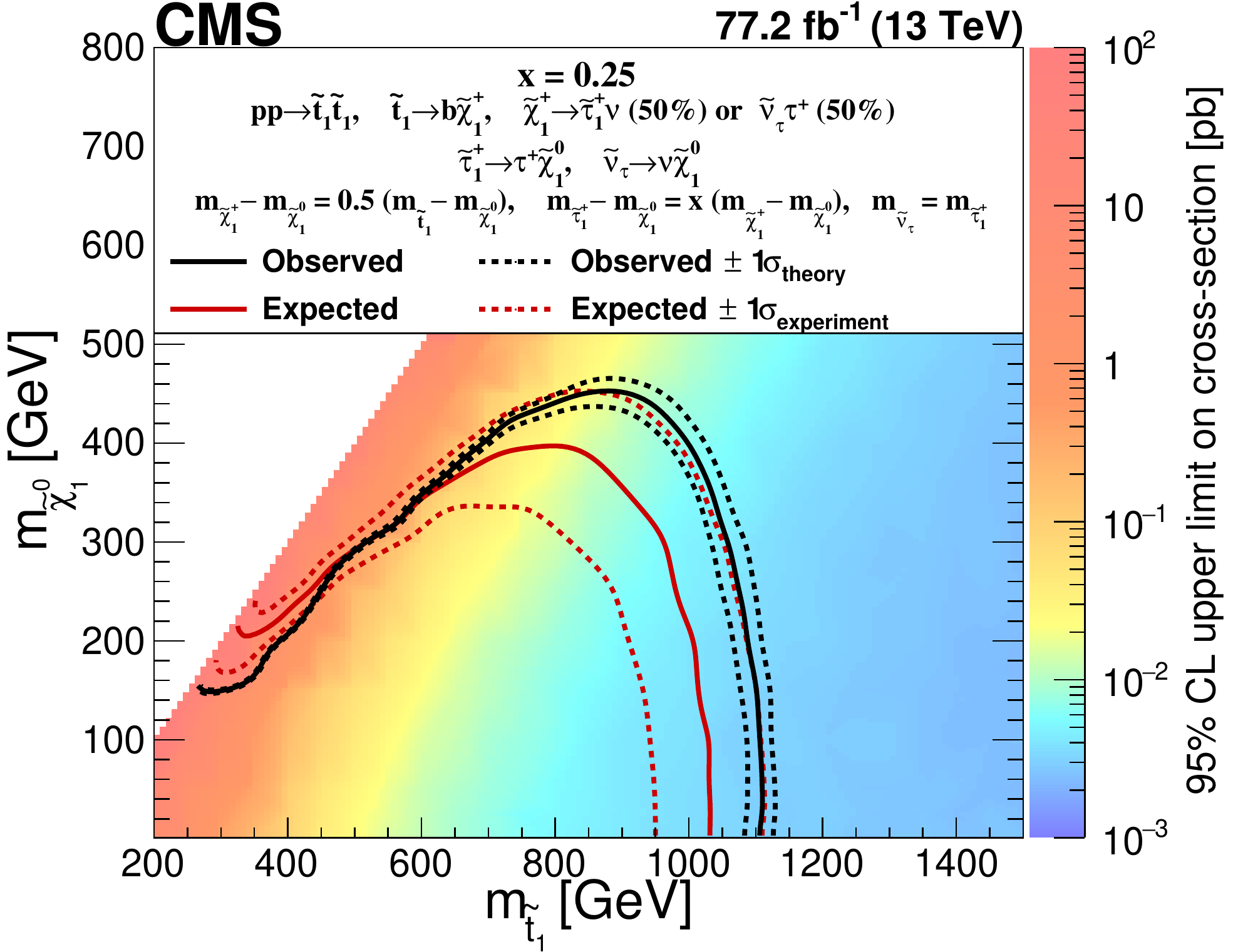}
	\includegraphics[width=0.495\textwidth]{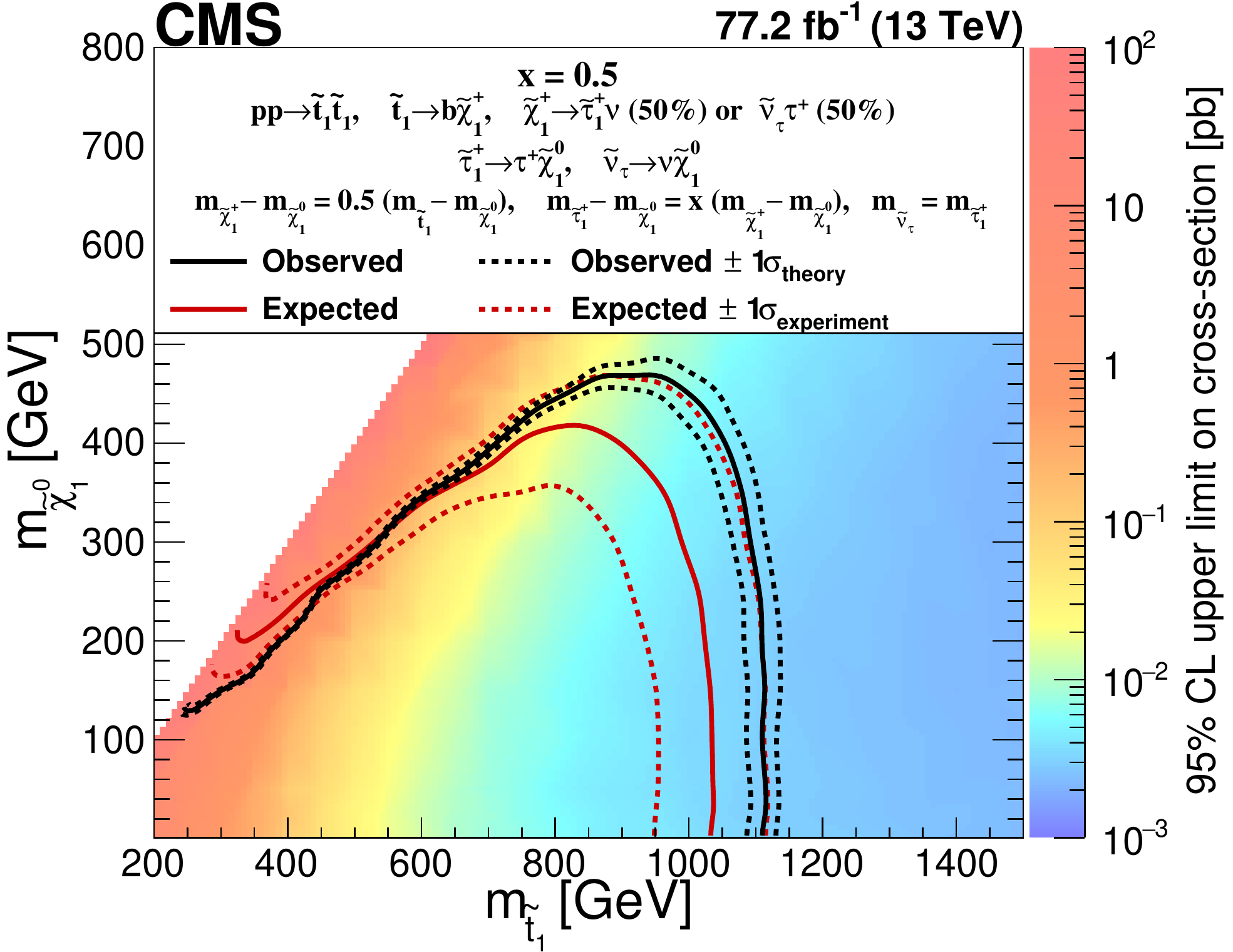} \\
	\includegraphics[width=0.495\textwidth]{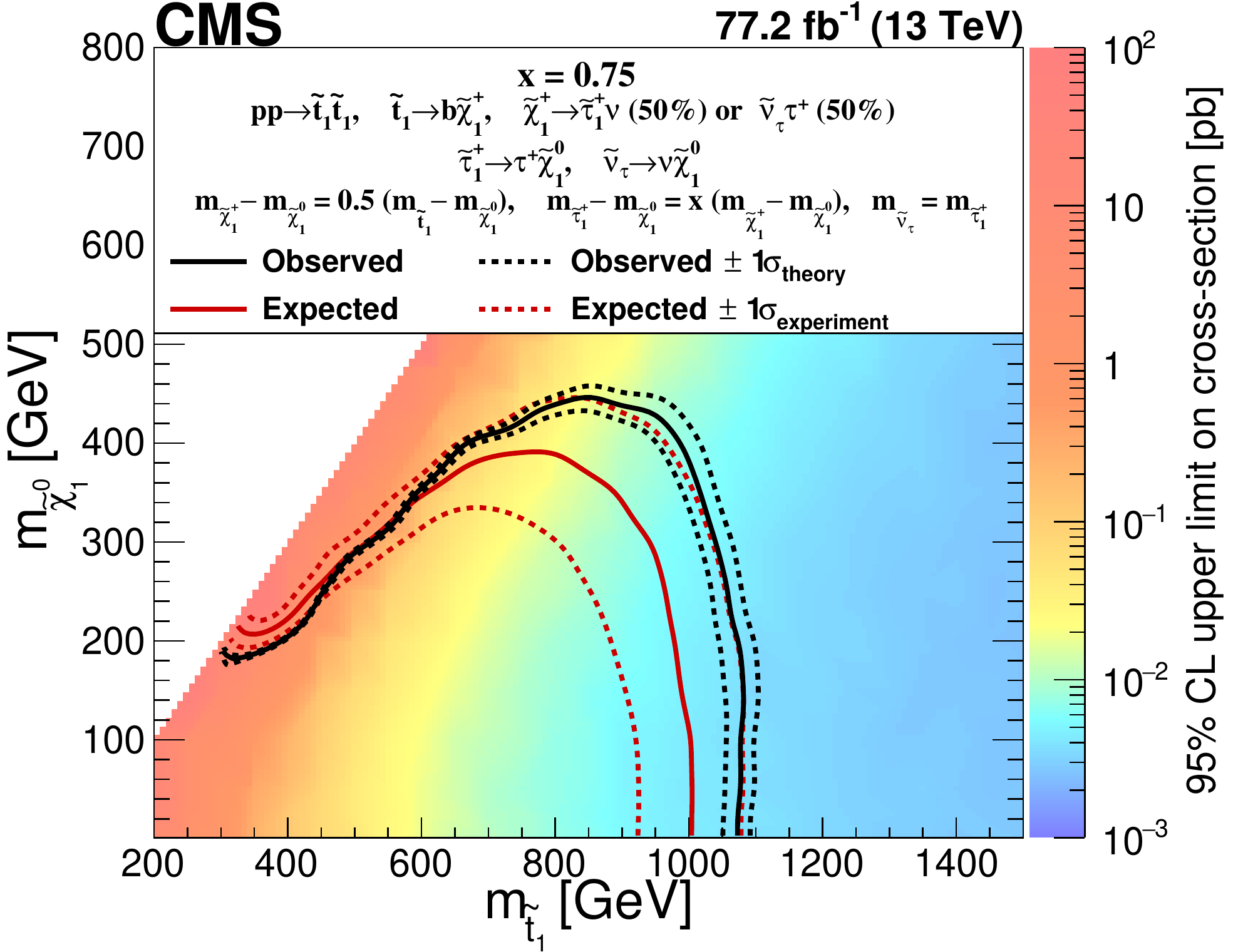}
	
	\caption{
		Exclusion limits at 95\% \CL for the pair production of top squarks decaying to a {\tauh}\tauh final state, displayed in the $ {m_{\PSQtDo}}$-$m_{\PSGczDo} $ plane for $ x =  0.25$ (upper left), 0.5 (upper right) and 0.75 (lower), as described in Eq.~(\ref{eq:mass}).
		The color axis represents the observed limit in the cross section, while the black (red) lines represent the observed (expected) mass limits.
		The signal cross sections are evaluated using NNLO plus next-to-leading logarithmic (NLL) calculations.
		The solid lines represent the central values.
		The dashed red lines indicate the region containing 68\% of the distribution of limits expected under the background-only hypothesis. The dashed black lines show the change in the observed limit due to variation of the signal cross sections within their theoretical uncertainties.
	}
	\label{fig:2Dlimit}
\end{figure}

Final results are obtained by combining the yields from 2016 and 2017 data sets.
The systematic uncertainties due to JES, factorization and
renormalization scales, misidentification rate measurement, and
\textsc{FastSim} \ptmiss correction are taken as correlated, and the rest of
the uncertainties are treated as uncorrelated between the two data sets.
The results are presented as observed and expected exclusion limits in
the top squark and LSP mass plane in Fig.~\ref{fig:2Dlimit}.
Top squark masses up to $ 1100\GeV$ are excluded for a nearly massless
LSP, and LSP masses up to $ 450\GeV$ are excluded for a top squark mass of $900\GeV$.
The exclusion limits are not very sensitive to the choice of $x$ because of the complementary nature of the signal diagrams, as discussed in Section \ref{sec:MCsimulation}.

The most sensitive search bins for the higher top squark masses are 14 and 15.
The observed data in these two bins are lower than the total background prediction,
resulting in the observed limit being higher than the expected one.
Hence, even though there are more events in data than prediction overall,
the observed mass limit is stronger than expected.
The excesses are
primarily in bins 2, 5, 7, and 12 which
are more significant for low top squark masses, hence
the observed limit is slightly worse than expected in that region.
The limits become weaker with decreasing $ \Delta m = m_{\PSQtDo} -
m_{\PSGczDo} $, corresponding to a parameter space with final-state particles having
lower momentum and hence less sensitivity.

\begin{table}[!htbp]
	\centering
	
	\topcaption{Event yields along with statistical and systematic
		uncertainties in the 2016 and 2017 analyses combined, for
		different background sources and the total background
		in the 15 search bins, as defined in Fig.~\ref{fig:sigBin}.
		The uncertainties that are smaller than 0.05 are listed as 0.0.
		The number of events observed in data is also shown.
		The notation used is $ \text{yield} ^{+\text{stat} +\text{syst}}_{-\text{stat}-\text{syst}} $.}
	
	\begin{tabular}{ l  c  c  c  c  c  c }
		\hline
		SR &
		$ \ttbar $ &
		DY+jets &
		Other SM &
		Misid. \tauh &
		Total bkg. &
		Data \\
		
		\hline
		\\
		1 &
		$ 170 ^{+8 +21} _{-8 -19} $ &
		$ 98 ^{+10 +22} _{-10 -20} $ &
		$ 23 ^{+2 +4} _{-2 -4} $ &
		$ 150 ^{+21 +6} _{-21 -19} $ &
		$ 441 ^{+25 +42} _{-25 -43} $ &
		$ 417 $ \\[\cmsTabSkip]
		
		2 &
		$ 200 ^{+8 +22} _{-8 -21} $ &
		$ 154 ^{+11 +35} _{-11 -33} $ &
		$ 29 ^{+3 +6} _{-3 -5} $ &
		$ 94 ^{+19 +22} _{-19 -20} $ &
		$ 477 ^{+23 +59} _{-23 -55} $ &
		$ 559 $ \\[\cmsTabSkip]
		
		3 &
		$ 20 ^{+3 +2} _{-3 -2} $ &
		$ 20 ^{+3 +6} _{-3 -5} $ &
		$ 5.1 ^{+0.8 +1.1} _{-0.8 -1.6} $ &
		$ 12 ^{+6 +4} _{-6 -3} $ &
		$ 57 ^{+7 +9} _{-7 -8} $ &
		$ 49 $ \\[\cmsTabSkip]
		
		4 &
		$ 19 ^{+3 +2} _{-3 -2} $ &
		$ 4.1 ^{+1.5 +1.0} _{-1.5 -1.0} $ &
		$ 3.8 ^{+0.8 +0.9} _{-0.8 -0.9} $ &
		$ 3.2 ^{+3.5 +1.2} _{-3.5 -0.9} $ &
		$ 30 ^{+5 +4} _{-5 -4} $ &
		$ 28 $ \\[\cmsTabSkip]
		
		5 &
		$ 8.9 ^{+1.6 +1.0} _{-1.6 -1.0} $ &
		$ 0.5 ^{+4.8 +0.1} _{-0.5 -0.1} $ &
		$ 1.6 ^{+0.4 +0.4} _{-0.4 -0.3} $ &
		$ 3.8 ^{+3.2 +1.2} _{-3.2 -0.8} $ &
		$ 15 ^{+6 +2} _{-4 -1} $ &
		$ 22 $ \\[\cmsTabSkip]
		
		6 &
		$ 51 ^{+4 +6} _{-4 -5} $ &
		$ 36 ^{+6 +9} _{-6 -10} $ &
		$ 7.8 ^{+1.3 +1.6} _{-1.3 -1.8} $ &
		$ 78 ^{+14 +12} _{-14 -13} $ &
		$ 173 ^{+15 +19} _{-15 -20} $ &
		$ 169 $ \\[\cmsTabSkip]
		
		7 &
		$ 46 ^{+4 +5} _{-4 -5} $ &
		$ 14 ^{+4 +6} _{-4 -4} $ &
		$ 5.2 ^{+0.8 +1.0} _{-0.8 -1.1} $ &
		$ 48 ^{+11 +12} _{-11 -10} $ &
		$ 113 ^{+12 +15} _{-12 -13} $ &
		$ 133 $ \\[\cmsTabSkip]
		
		8 &
		$ 4.4 ^{+1.3 +0.6} _{-1.3 -0.5} $ &
		$ 4.3 ^{+1.5 +2.0} _{-1.5 -1.1} $ &
		$ 1.4 ^{+0.4 +0.3} _{-0.4 -0.3} $ &
		$ 9.0 ^{+3.2 +3.9} _{-3.2 -2.7} $ &
		$ 19 ^{+4 +5} _{-4 -3} $ &
		$ 23 $ \\[\cmsTabSkip]
		
		9 &
		$ 3.7 ^{+1.1 +0.5} _{-1.1 -0.4} $ &
		$ 0.0 ^{+3.5 +0.0} _{-0.0 -0.0} $ &
		$ 0.7 ^{+0.3 +0.2} _{-0.3 -0.2} $ &
		$ 4.6 ^{+1.7 +2.0} _{-1.7 -1.4} $ &
		$ 9.0 ^{+4.0 +2.1} _{-2.0 -1.5} $ &
		$ 4 $ \\[\cmsTabSkip]
		
		10 &
		$ 1.0 ^{+0.6 +0.2} _{-0.6 -0.1} $ &
		$ 0.0 ^{+3.5 +0.0} _{-0.0 -0.0} $ &
		$ 0.5 ^{+0.2 +0.1} _{-0.2 -0.1} $ &
		$ 3.2 ^{+1.4 +1.3} _{-1.4 -1.0} $ &
		$ 4.7 ^{+3.8 +1.3} _{-1.6 -1.0} $ &
		$ 1 $ \\[\cmsTabSkip]
		
		11 &
		$ 6.8 ^{+1.6 +1.2} _{-1.6 -0.9} $ &
		$ 2.4 ^{+1.5 +1.3} _{-1.5 -0.7} $ &
		$ 1.4 ^{+0.4 +0.7} _{-0.4 -0.3} $ &
		$ 16 ^{+6 +4} _{-6 -3} $ &
		$ 27 ^{+6 +5} _{-6 -4} $ &
		$ 30 $ \\[\cmsTabSkip]
		
		12 &
		$ 2.9 ^{+1.0 +0.4} _{-1.0 -0.3} $ &
		$ 8.3 ^{+2.5 +2.8} _{-2.5 -1.8} $ &
		$ 2.1 ^{+0.3 +0.4} _{-0.3 -0.4} $ &
		$ 11 ^{+6 +2} _{-6 -2} $ &
		$ 24 ^{+7 +4} _{-7 -3} $ &
		$ 41 $ \\[\cmsTabSkip]
		
		13 &
		$ 0.7 ^{+0.9 +0.2} _{-0.5 -0.4} $ &
		$ 2.2 ^{+0.9 +0.6} _{-0.9 -0.9} $ &
		$ 0.7 ^{+0.2 +0.2} _{-0.2 -0.2} $ &
		$ 3.5 ^{+2.1 +1.3} _{-2.1 -0.9} $ &
		$ 7.1 ^{+2.4 +1.5} _{-2.3 -1.4} $ &
		$ 6 $ \\[\cmsTabSkip]
		
		14 &
		$ 0.5 ^{+0.9 +0.1} _{-0.5 -0.1} $ &
		$ 0.0 ^{+3.5 +0.0} _{-0.0 -0.0} $ &
		$ 0.6 ^{+0.2 +0.1} _{-0.2 -0.1} $ &
		$ 0.0 ^{+1.8 +0.0} _{-0.0 -0.0} $ &
		$ 1.1 ^{+4.1 +0.2} _{-0.6 -0.2} $ &
		$ 1 $ \\[\cmsTabSkip]
		
		15 &
		$ 0.3 ^{+0.8 +0.1} _{-0.3 -0.1} $ &
		$ 0.4 ^{+2.2 +0.1} _{-0.4 -0.1} $ &
		$ 0.1 ^{+0.0 +0.0} _{-0.0 -0.0} $ &
		$ 1.0 ^{+1.8 +0.4} _{-0.0 -0.3} $ &
		$ 1.8 ^{+3.0 +0.5} _{-0.5 -0.3} $ &
		$ 0 $ \\[\cmsTabSkip]
		
		Total &
		$ 535 ^{+14 +63} _{-14 -58} $ &
		$ 344 ^{+19 +78} _{-17 -73} $ &
		$ 83 ^{+4 +16} _{-4 -16} $ &
		$ 437 ^{+35 +73} _{-35 -78} $ &
		$ 1400 ^{+43 +125} _{-42 -123} $ &
		$ 1483 $ \\[\cmsTabSkip]
		
		\hline
	\end{tabular}
	
	\label{tab:Bg20162017}
\end{table}

\section{Summary}
\label{sec:Summary}

The signature of top squark pair production in final states with two tau leptons has been explored in data collected with the CMS
detector during 2016 and 2017, corresponding to integrated luminosities of
35.9  and 41.3\fbinv, respectively.
The search was performed in the final state containing an oppositely charged hadronic tau lepton pair, at least one jet identified as likely to originate from the fragmentation of a \PQb quark, and missing transverse momentum.
The dominant standard model backgrounds were found to originate from top quark pair production and processes where jets were misidentified as hadronic tau lepton decays. Control samples in data were used to estimate these backgrounds, while
other backgrounds were estimated using simulation.

No significant excess was observed, and exclusion limits on the top squark mass in terms of the mass of the lightest neutralino were set at 95\% confidence level within the framework of simplified models where the top squark decays via a chargino to final states including tau leptons.
In such models, top squark masses are excluded up to 1100\GeV for an almost massless neutralino, and LSP masses up to 450\GeV are excluded for a top squark mass of 900\GeV.
These results probe a region of the supersymmetric parameter space corresponding to high-\tanb and higgsino-like scenarios.

\begin{acknowledgments}
We congratulate our colleagues in the CERN accelerator departments for the excellent performance of the LHC and thank the technical and administrative staffs at CERN and at other CMS institutes for their contributions to the success of the CMS effort. In addition, we gratefully acknowledge the computing centers and personnel of the Worldwide LHC Computing Grid for delivering so effectively the computing infrastructure essential to our analyses. Finally, we acknowledge the enduring support for the construction and operation of the LHC and the CMS detector provided by the following funding agencies: BMBWF and FWF (Austria); FNRS and FWO (Belgium); CNPq, CAPES, FAPERJ, FAPERGS, and FAPESP (Brazil); MES (Bulgaria); CERN; CAS, MoST, and NSFC (China); COLCIENCIAS (Colombia); MSES and CSF (Croatia); RPF (Cyprus); SENESCYT (Ecuador); MoER, ERC IUT, PUT and ERDF (Estonia); Academy of Finland, MEC, and HIP (Finland); CEA and CNRS/IN2P3 (France); BMBF, DFG, and HGF (Germany); GSRT (Greece); NKFIA (Hungary); DAE and DST (India); IPM (Iran); SFI (Ireland); INFN (Italy); MSIP and NRF (Republic of Korea); MES (Latvia); LAS (Lithuania); MOE and UM (Malaysia); BUAP, CINVESTAV, CONACYT, LNS, SEP, and UASLP-FAI (Mexico); MOS (Montenegro); MBIE (New Zealand); PAEC (Pakistan); MSHE and NSC (Poland); FCT (Portugal); JINR (Dubna); MON, RosAtom, RAS, RFBR, and NRC KI (Russia); MESTD (Serbia); SEIDI, CPAN, PCTI, and FEDER (Spain); MOSTR (Sri Lanka); Swiss Funding Agencies (Switzerland); MST (Taipei); ThEPCenter, IPST, STAR, and NSTDA (Thailand); TUBITAK and TAEK (Turkey); NASU (Ukraine); STFC (United Kingdom); DOE and NSF (USA).

\hyphenation{Rachada-pisek} Individuals have received support from the Marie-Curie program and the European Research Council and Horizon 2020 Grant, contract Nos.\ 675440, 752730, and 765710 (European Union); the Leventis Foundation; the A.P.\ Sloan Foundation; the Alexander von Humboldt Foundation; the Belgian Federal Science Policy Office; the Fonds pour la Formation \`a la Recherche dans l'Industrie et dans l'Agriculture (FRIA-Belgium); the Agentschap voor Innovatie door Wetenschap en Technologie (IWT-Belgium); the F.R.S.-FNRS and FWO (Belgium) under the ``Excellence of Science -- EOS" -- be.h project n.\ 30820817; the Beijing Municipal Science \& Technology Commission, No. Z181100004218003; the Ministry of Education, Youth and Sports (MEYS) of the Czech Republic; the Lend\"ulet (``Momentum") Program and the J\'anos Bolyai Research Scholarship of the Hungarian Academy of Sciences, the New National Excellence Program \'UNKP, the NKFIA research grants 123842, 123959, 124845, 124850, 125105, 128713, 128786, and 129058 (Hungary); the Council of Science and Industrial Research, India; the HOMING PLUS program of the Foundation for Polish Science, cofinanced from European Union, Regional Development Fund, the Mobility Plus program of the Ministry of Science and Higher Education, the National Science Center (Poland), contracts Harmonia 2014/14/M/ST2/00428, Opus 2014/13/B/ST2/02543, 2014/15/B/ST2/03998, and 2015/19/B/ST2/02861, Sonata-bis 2012/07/E/ST2/01406; the National Priorities Research Program by Qatar National Research Fund; the Ministry of Science and Education, grant no. 3.2989.2017 (Russia); the Programa Estatal de Fomento de la Investigaci{\'o}n Cient{\'i}fica y T{\'e}cnica de Excelencia Mar\'{\i}a de Maeztu, grant MDM-2015-0509 and the Programa Severo Ochoa del Principado de Asturias; the Thalis and Aristeia programs cofinanced by EU-ESF and the Greek NSRF; the Rachadapisek Sompot Fund for Postdoctoral Fellowship, Chulalongkorn University and the Chulalongkorn Academic into Its 2nd Century Project Advancement Project (Thailand); the Nvidia Corporation; the Welch Foundation, contract C-1845; and the Weston Havens Foundation (USA). 	
\end{acknowledgments}

\bibliography{auto_generated}

\cleardoublepage \appendix\section{The CMS Collaboration \label{app:collab}}\begin{sloppypar}\hyphenpenalty=5000\widowpenalty=500\clubpenalty=5000\vskip\cmsinstskip
\textbf{Yerevan Physics Institute, Yerevan, Armenia}\\*[0pt]
A.M.~Sirunyan$^{\textrm{\dag}}$, A.~Tumasyan
\vskip\cmsinstskip
\textbf{Institut f\"{u}r Hochenergiephysik, Wien, Austria}\\*[0pt]
W.~Adam, F.~Ambrogi, T.~Bergauer, J.~Brandstetter, M.~Dragicevic, J.~Er\"{o}, A.~Escalante~Del~Valle, M.~Flechl, R.~Fr\"{u}hwirth\cmsAuthorMark{1}, M.~Jeitler\cmsAuthorMark{1}, N.~Krammer, I.~Kr\"{a}tschmer, D.~Liko, T.~Madlener, I.~Mikulec, N.~Rad, J.~Schieck\cmsAuthorMark{1}, R.~Sch\"{o}fbeck, M.~Spanring, D.~Spitzbart, W.~Waltenberger, C.-E.~Wulz\cmsAuthorMark{1}, M.~Zarucki
% ...
\vskip\cmsinstskip
\dag: Deceased\\
1:  Also at Vienna University of Technology, Vienna, Austria\\
% ...
\end{sloppypar}
\end{document}